\documentclass[a4paper,12pt,english]{article}

\usepackage{latexsym,epsfig,amssymb, amsmath, cite}%, nicefrac}
\usepackage{amsmath}

\newcommand{\bea}{\setlength\arraycolsep{2pt} \begin{eqnarray}}
\newcommand{\eea}{\end{eqnarray}}

\usepackage{hyperref}

\newsavebox{\uuunit}
\sbox{\uuunit}
{\setlength{\unitlength}{0.825em}
	\begin{picture}(0.6,0.7)
	\thinlines
	\put(0,0){\line(1,0){0.5}}
	\put(0.15,0){\line(0,1){0.7}}
	\put(0.35,0){\line(0,1){0.8}}
	\multiput(0.3,0.8)(-0.04,-0.02){12}{\rule{0.5pt}{0.5pt}}
	\end {picture}}

\def\be{\begin{equation}}
\def\ee{\end{equation}}
\def\ba{\begin{array}}
\def\ea{\end{array}}
\def\bea{\begin{eqnarray}}
\def\eea{\end{eqnarray}}
\def\bd{\begin{displaymath}}
\def\ed{\end{displaymath}}

\def\a{\alpha}
\def\b{\beta}

\def\f{\phi}

\def\l{\lambda}
\def\L{\Lambda}
\def\m{\mu}
\def\n{\nu}
\def\r{\rho}
\def\s{\sigma}

\def\o{\omega}

\newcommand{\de}{\partial}
\newcommand{\no}{\noindent}
\newcommand{\nb}{\nonumber}
\newcommand\Ost{Ostrogradsky }

\def\ll{\mathcal{L}}
\def\Q{\mathcal{Q}}
\def\E{\mathcal{E}}

\begin{document}

\makeatother

\parindent=0cm

\renewcommand{\title}[1]{\vspace{10mm}\noindent{\Large{\bf

#1}}\vspace{8mm}} \newcommand{\authors}[1]{\noindent{\large

#1}\vspace{5mm}} \newcommand{\address}[1]{{\itshape #1\vspace{2mm}}}

\begin{titlepage}

\begin{flushright}
\today \\
\end{flushright}

\begin{center}

\vskip 3mm

\title{ {\LARGE
Higher Derivative Field Theories:  \\ \medskip Degeneracy Conditions and Classes 
} }

 \authors{Marco Crisostomi$^{\a}$,  Remko Klein$^{\b}$ and Diederik Roest$^{\b}$
}

\vskip 1mm

\address{
$^{\a}$Institute of Cosmology and Gravitation, University of Portsmouth, \\ Portsmouth, PO1 3FX, UK

\smallskip 

$^{\beta}$Van Swinderen Institute for Particle Physics and 
Gravity, University of Groningen, \\ 
Nijenborgh 4, 9747 AG Groningen, The Netherlands

}

\smallskip

\verb+ marco.crisostomi@port.ac.uk, remko.klein@rug.nl, d.roest@rug.nl +

\end{center}

\vskip 1mm

 \begin{center}
%%%% --- ABSTRACT --- %%%%
\textbf{Abstract}
\vskip 3mm
\begin{minipage}{14cm}%/
We provide a full analysis of ghost free higher derivative field theories with coupled degrees of freedom. Assuming the absence of gauge symmetries, we derive the degeneracy conditions in order to evade the Ostrogradsky ghosts, and analyze which (non)trivial classes of solutions this allows for. It is shown explicitly how Lorentz invariance avoids the propagation of ``half'' degrees of freedom. Moreover, for a large class of theories, we construct the field redefinitions and/or (extended) contact transformations that put the theory in a manifestly first order form. Finally, we identify which class of theories cannot be brought to first order form by such transformations.

\end{minipage}

 \end{center}

\end{titlepage}

\tableofcontents

%\newpage

\setcounter{footnote}{0}

\section{Introduction}

\no The last few years have seen a growing interest in higher derivative theories, i.e.~theories with second or higher derivatives in the action, mainly motivated by model building for inflation and dark energy or modified gravity. Much of this work builds on the old theorem of Ostrogradsky \cite{Ostrogradski}. This theorem implies that, in the absence of any degeneracies\footnote{The word ``degenerate'' is associated with the Hessian matrix of the Lagrangian with respect to the velocities. A degenerate Hessian matrix implies that the system of momenta cannot be inverted and therefore there are primary constraints. {Further details can be found in Appendices A and B as well as e.g.~\cite{Algorithm1, Algorithm2}.}}, a higher derivative theory will have additional degrees of freedom that are ghost like. Therefore, healthy higher derivative theories are necessarily degenerate, i.e.~they are constrained systems.

In the simple example of a mechanical system with a single variable, it can be seen that any degenerate higher derivative theory amounts to an ordinary and thus healthy theory, with at most first derivatives in the action, up to an irrelevant total derivative. Such higher derivative theories are therefore trivial. This invites question marks regarding the usefulness of higher derivative theories. Interestingly, in more general contexts, there can be degenerate yet nontrivial higher derivative theories.

The first step beyond trivial higher derivatives regards {\it field theories}. A prime example is (generalized) Galileon theories, consisting of a single scalar field with Lorentz invariant higher derivative interactions \cite{Nicolis:2008in, Deffayet:2009mn}. A similar construction for the spin-2 tensor leads to Lovelock gravity with specific $R^n$ interactions  \cite{Lovelock:1971yv}. Remarkably, these interactions have been chosen such that they still lead to second order field equations, thus evading the Ostrogradsky theorem\footnote{Note that having second order field equations in a degenerate theory does not guarantee the absence of additional ghosts; in general other conditions might be necessary. In fact, in some cases these additional ghosts are actually interpretable as Ostrogradsky ghosts upon using a different field basis to describe the theory, see e.g.~massive gravity \cite{deRham:2010kj, deRham:2011rn} and vector Galileons \cite{Heisenberg:2014rta}.}. This can be understood by the observation that the higher derivative interactions can be packaged into an ordinary Lagrangian plus a total derivative, similar to the mechanics case; however, this ordinary Lagrangian cannot be written in  a manifestly Lorentz invariant form. This trade off between manifest first order Lagrangians and manifestly Lorentz invariant Lagrangians (and the impossibility to have both) will be a recurring theme in the present paper. 

A second generalization, and equally relevant to the present work, concerns {\it coupled systems} with multiple variables or fields. Similar to the case with a single variable, for many years the community only trusted a very special subset of these theories, namely the ones giving second order field equations while (erroneously) assuming that all the others are plagued by instabilities. For instance, the most general scalar-tensor theories with second order field equations are those of Horndeski \cite{Horndeski:1974wa}, which coincide \cite{Kobayashi:2011nu} with covariantized generalized Galileons \cite{Deffayet:2011gz,Deffayet:2009wt}. Similarly,  covariant vector Galileons describe such couplings between a vector and tensor \cite{Tasinato:2014eka, Heisenberg:2014rta, Hull:2015uwa}. Very recently this was generalized to covariant tensor Galileons for the couplings between different tensors \cite{Thanasis}.

Only very recently it has been realised that one can have healthy degenerate higher derivative theories even in the presence of higher order field equations, with the proposal of beyond Horndeski models \cite{Zumalacarregui:2013pma, Gleyzes:2014dya, Gleyzes:2014qga}. These models have been further understood and generalised in \cite{Deffayet:2015qwa, Langlois:2015cwa, Langlois:2015skt, Crisostomi:2016tcp, Crisostomi:2016czh, Achour:2016rkg, deRham:2016wji, Ezquiaga:2017ner} and now a complete classification for degenerate scalar-tensor theories within a certain Ansatz exists \cite{BenAchour:2016fzp}.
{Analogously, similar constructions for vector interactions were introduced in \cite{Heisenberg:2016eld} and a classification for degenerate vector-tensor theories (up to quadratic order) was given in \cite{Kimura:2016rzw}.}

A central theme of these constructions is the coupling between a higher derivative degree of freedom and a healthy first order one. In the above examples, these are a scalar and a tensor or a vector and a tensor, respectively. This interplay between higher derivative and healthy sectors was analyzed in full generality in the mechanics case in \cite{Motohashi:2016ftl} and \cite{Klein:2016aiq}, where respectively a Hamiltonian and Lagrangian constraint analysis have been performed, leading to conditions that ensure the absence of Ostrogradsky ghosts. The aim of this paper is to perform a similarly general analysis for the case of (Lorentz invariant) field theories, and to classify which nontrivial theories this allows for. 

Specifically, we consider field theories whose Lagrangians depend on $M$ higher derivative fields $\f_m$ and $A$ 'healthy' fields $q_\a$:
\begin{align} 
L(\partial_\m\partial_\n \f_m, \partial_\m \f_m, \f_m, \partial_\m q_\a, q_\a) \,. \label{Lagstudied}
\end{align}
We only include dependence up to second derivatives\footnote{Note that we do not include dependence on mixed or pure spatial higher order derivatives,  e.g.~$\partial_i \dot{q}_\alpha$, $\partial_i \ddot{\f}_m$, $\partial_i\partial_j q_\a$, etc. which would be allowed in non-Lorentz invariant field theories. Although including such dependences would in principle modify the analysis and the resulting degeneracy conditions, we believe the general structures will remain unchanged. Therefore, in order to not clutter up the formulae and the discussion, we refrain from this more general analysis.}; we will comment on yet higher derivative structures in the concluding section. Moreover, we make the following assumptions:
\begin{itemize}
\item The higher derivative fields are treated on an equal footing in the sense that we assume all the constraints coming in sets of $M$.  Since we are only interested in the case where the Lagrangians truly depend on the second order time derivatives of the higher derivative fields, we assume that\footnote{Throughout the paper we use the notation $L_\psi \equiv \de L / \de \psi$, where $\psi$ can be a field or space/time derivatives of a field. Later we will also use the notation $E_\psi$ to denote the equations of motion respect to $\psi$.} $L_{\ddot{\phi}_m} \neq 0$ for all $m$.
Also,  we aim to remove only the \Ost modes, so we do not consider the case of extra constraints that further reduce the number of degrees of freedom (dof).
\item The theories we consider posses no gauge symmetries. In the Lagrangian analysis this means that we do not encounter any gauge identities, i.e. combinations of equations of motion (eom) that vanish identically. In the Hamiltonian analysis this means that no first class constraints are present, i.e. we assume all constraints to be second class.
\item We are not interested in possible degeneracies in the healthy sector. We thus assume that the healthy sector itself is non-degenerate, which is precisely the case when the kinetic matrix $L_{\dot{q}_\a\dot{q}_\b}$ is invertible.
\end{itemize}

No further assumptions are made about the functional dependence of the Lagrangian; f.e. it does not need to be polynomial in the highest derivatives. Also, we do not assume any global symmetry, space-time or internal. This means that we also consider Lorentz violating theories, although we will also specifically address Lorentz invariant ones.

Let us conclude by giving a short overview of the structure of the paper. In Section \ref{struccond} we state (the complete analyses can be found in Appendix \ref{Laganalysis} and \ref{Hamanalysis}) and interpret our results following from the Lagrangian and Hamiltonian analyses of the theories described above. Specifically, we analyse the conditions to remove \Ost modes, in particular in relation to the structure of the eom and the counting of dof. We first review the results already obtained for mechanical systems and subsequently generalise them to the field theory case. We conclude the section with a discussion of the special properties of Lorentz invariant theories. In Section \ref{classification} we propose a formal classification for healthy higher derivative theories and analyse their properties in more detail. In particular we discuss how different classes can be related via field redefinitions and/or (extended) contact transformations. Again we give special attention to Lorentz invariant theories. We draw a number of conclusions in Section \ref{conclusions}.

\section{Structure of degeneracy conditions}
\label{struccond}

\no In this section we analyze and discuss the degeneracy conditions, and their implication for the field equations, for three different systems: mechanics and (Lorentz invariant) field theories. The field theory derivation closely follows that of mechanics, which has been performed both in the Hamiltonian  \cite{Motohashi:2016ftl} and Lagrangian \cite{Klein:2016aiq} framework, and therefore has been placed in Appendix \ref{Laganalysis} and \ref{Hamanalysis}. 

\subsection{Mechanical systems}

\no We will start with a short recap of the results of \cite{Motohashi:2016ftl, Klein:2016aiq}. Starting from a generic Lagrangian
\begin{align} 
L = L(\ddot{\f}_m,\dot{\f}_m,\f_m,\dot{q}_\a,q_\a) \,,
\end{align} 
that satisfies the assumptions in the Introduction, one can put the theory in a first order form using auxiliary fields, and perform a Lagrangian and/or Hamiltonian analysis to determine the number of propagating dof. For a generic theory, i.e. non-degenerate, it follows that no constraints are present and the theory propagates $2M + A$ degrees of freedom, $M$ of which are Ostrogradsky ghosts. Healthy theories are therefore necessarily  degenerate (constrained) systems. 

A key concept in the discussion of the degeneracy conditions are the vectors
 \begin{align}
  v_m^A = (\delta_m^n , V_m^\a) \quad \text{~~with~~} \quad V_m^\a \equiv -L_{\ddot{\phi}_m\dot{q}_\b}L^{-1}_{\dot{q}_\b\dot{q}_\a} \,, \label{eigenvectors}
 \end{align}
 where the index $A$ spans over the set $(n,\a)$.
%These span $M$ orthogonal directions.
The {\it primary} conditions amount to the requirement that these are null eigenvectors\footnote{Due to the normalization used in (\ref{eigenvectors}), in the following we will often refer to the components $V_m^\a$ as the null eigenvectors themselves.} of the Hessian matrix of the Lagrangian with respect to the velocities $\dot{\psi}_A$ of the collection $\psi_A \equiv (\dot{\f}_m, q_\a)$:
 \begin{align}
0 =  P_{(mn)}  &\equiv v^A_m L_{\dot{\psi}_A \dot{\psi}_B}v^B_n  \nb \\
&= L_{\ddot{\f}_m \ddot{\f}_n} + L_{\ddot{\f}_m\dot{q}_\a}V^\a_n  \,. 
 \end{align}
 Additionally, one must satisfy the \textit{secondary} conditions:
\begin{align}  
0 = S_{[mn]} &\equiv 2\, v^A_m L_{\dot{\psi}_{[A} \psi_{B]}}v^B_n \nb \\
&= 2\left(L_{\ddot{\f}_{[m} \dot{\f}_{n]}}  + V_{[m}^\a L_{\dot{q}_\a \dot{\f}_{n]}} + L_{\ddot{\f}_{[m} q_\b} V_{n]}^\b  + V_m^\a L_{\dot{q}_{[\a} q_{\b]}} V_n^\b \right) \,.
\end{align} 
Satisfying the primary conditions  ensures the existence of $M$ primary constraints.
%equations / primary  constraints in the Lagrangian / Hamiltonian analyses.
The secondary conditions   in turn guarantee the existence of $M$ secondary constraints.
%Lagrangian constraint equations / Hamiltonian secondary constraints.
Therefore, if one satisfies both, a total number of $2M$ constraints
%Lagrangian constraint equations / Hamiltonian constraints
are present and we end up with a total of $2M + A - \frac{1}{2}(2M) = M + A$ dof.
All the Ostrogradsky degrees of freedom are absent.

\smallskip

The role of the primary and secondary conditions can be made clear at the level of the original equations of motion. First observe that one can always, whether the conditions are satisfied or not, get rid of the third and second order time derivatives of $q_\a$ in $E_{\f_m}$ by considering the combination:
\begin{align}
E_{\f_m} + \frac{d}{dt}(V^\a_m E_{q_\a}) + U^\a_m E_{q_\a} = P_{(mn)}\f^{(4)}_n + \left( \dddot{\f}, \dot{q}, \dots \right) \,,
\end{align}
where $U^\a_m$ is defined in (\ref{U}).
If the primary and secondary conditions are not satisfied, this is the best one can do. One can in principle solve for $\f_m$ if one specifies $4M+2A$ initial conditions, $(\dddot{\f}_m,\ddot{\f}_m,\dot{\f}_m,\f_m)_0$ and $(\dot{q}_\a,q_\a)_0$. Since $E_{q_\a}$ depends on at most $\dddot{\f}$ and $\ddot{q}$, one can subsequently solve for $\ddot{q}_\a$ without having to specify additional initial conditions. Hence $\frac{1}{2}(4M+2A) = 2M + A$ dof propagate.

On the other hand, if the primary conditions are satisfied, the $\f^{(4)}_m$ terms and also the terms nonlinear in $\dddot{\f}_m$ are absent and one finds
\begin{align}
E_{\f_m} + \frac{d}{dt}(V^\a_m E_{q_\a}) + U^\a_m E_{q_\a} = S_{[mn]}\dddot{\f}_n +  \left( \ddot{\f}, \dot{q}, \dots \right) \,.
\end{align}
If also the secondary conditions hold, the terms linear in $\dddot{\f}_m$ drop out and one ends up with equations that contain at most $\ddot{\f}_m$ and $\dot{q}_\a$. These particular combinations thus tell us that one can express the initial values  $(\dddot{\f}_m,\ddot{\f}_m)_0$ in terms of $(\dot{\f}_m,\f_m)_0$ and $(\dot{q}_\a,q_\a)_0$. Therefore, to solve the full set of equations of motion, one only needs to specify $2M + 2A$ initial conditions, implying that   $M + A$ degrees of freedom propagate.

\smallskip

Let us conclude the discussion observing that $P_{(mn)}$ and $S_{[mn]}$ are generically independent; indeed, there exist theories where the primary conditions are satisfied but the secondary are not. Let us see what this structure implies for the number of degrees of freedom of such theories. First assume that we have an even number of primary constraints. Generically no secondary constraints are present and one finds an integer number of degrees of freedom. Now assume that there are an odd number of primary constraints. In this case there is automatically also 1 secondary constraint: since $S_{[mn]}$ is antisymmetric and odd-dimensional, it has one null eigenvalue, leading therefore to a secondary constraint. Thus also in the case of an odd number of primary constraints, one generically has an even number of total constraints and so an integer number of degrees of freedom.
Let us note however that these partially degenerate theories are still haunted by Ostrogradsky ghosts unless the secondary constraint is complemented by additional (tertiary, quartic, etc.) ones \cite{Motohashi:2014opa}. 
Note that the antisymmetry of the secondary conditions implies that if only one higher derivative variable is present, the primary condition actually implies the secondary condition.

\subsection{Field theories}
\label{Field theories}

\no Now, let us look at the analysis for the field theory case. Starting from 
\begin{align} 
L(\partial_\m\partial_\n \f_m, \partial_\m \f_m, \f_m, \partial_\m q_\a, q_\a) \,,
\end{align}
again one can put the Lagrangian in a first order form via the introduction of auxiliary fields and perform a Lagrangian and Hamiltonian constraint analysis. We have performed both the analyses whose details are given in Appendix \ref{Laganalysis} and \ref{Hamanalysis}.

In particular we find that in order to eliminate the Ostrogradsky modes one must now satisfy three sets of conditions, namely one set of primary conditions and two sets of secondary conditions:  % As noted this can be easily done if by simply using that $\bar{L}_{\partial_i A_m} =2 L_{\partial_i\dot{\f}_m}$. Doing this we end up with our final conditions:
\begin{align}
0 = P_{(mn)} &\equiv v^A_m L_{\dot{\psi}_A \dot{\psi}_B}v^B_n \nb \\
&= L_{\ddot{\f}_m \ddot{\f}_n} + L_{\ddot{\f}_m\dot{q}_\a}V^\alpha_n  \,, \label{primarycond}  \\[1ex]
0 = (S_i)_{(mn)} &\equiv 2\, v^A_m L_{\dot{\psi}_{(A} \de_i \psi_{B)}}v^B_n \nb \\
&=  2L_{\ddot{\f}_{(m}\partial_i \dot{\f}_{n)}} + 2V_{(m}^\a \left(L_{\dot{q}_\a \partial_i \dot{\f}_{n)}} + L_{ \partial_i q_\a \ddot{\f}_{n)}}\right)  + 2V_m^\a L_{\dot{q}_{(\a}\partial_i q_{\b)}}V^\b_n \,, \displaybreak[3]  \label{symmetricsecondary} \\[1ex]
0 = S_{[mn]} &\equiv  2\, v^A_m L_{\dot{\psi}_{[A} \psi_{B]}}v^B_n \nb + 2\, v^A_{[m} L_{\dot{\psi}_A \de_i \psi_{B}}\de_i v^B_{n]}  -  \de_i \left( v^A_m L_{\dot{\psi}_{[A} \de_i \psi_{B]}}v^B_n \right) \\
& = 2\left(L_{\ddot{\f}_{[m} \dot{\f}_{n]}}  + V_{[m}^\a L_{\dot{q}_\a \dot{\f}_{n]}} + L_{\ddot{\f}_{[m} q_\b} V_{n]}^\b  + V_m^\a L_{\dot{q}_{[\a} q_{\b]}} V_n^\b \right) \nb \\
& + \partial_i L_{\partial_i \dot{\f}_{[m} \ddot{\f}_{n]}}  + V_{[m}^\a \partial_i L_{\partial_i q_\a \ddot{\f}_{n]}} + \partial_i L_{\partial_i \dot{\f}_{[m} \dot{q}_\b} V_{n]}^\b + V_m^\a \partial_i L_{\partial_i q_{[\a} \dot{q}_{\b]}}V_n^\b \nb  \\
&+ \partial_i V^\b_{[n} \left( L_{\partial_i \dot{\f}_{m]} \dot{q}_\b}+ L_{\ddot{\f}_{m]} \partial_i q_\b } + 2V_{m]}^\a L_{\dot{q}_{(\a}\partial_i q_{\b)}}\right)  \,. \label{secondarycond}
\end{align}
Similarly to the mechanics case, satisfying the primary conditions enforces the existence of $M$ primary constraints. In order to have also $M$ secondary constraints, one must now satisfy both the secondary conditions.

\smallskip

Again the role of the conditions becomes clear when looking at the equations of motion. Regardless of whether one satisfies any of the constraints, one can always get rid of $\dddot{q}_\a$, $\partial_i \ddot{q}_\a$ and $\ddot{q}_\a$ in $E_{\f_m}$, by considering the following combination of equations
\begin{align} 
E_{\f_m} + \frac{d}{dt}(V^\a_m E_{q_\a}) + \partial_i (\a^{i\a}_m E_{q_\a}) + U^\a_m E_{q_\a} = P_{(mn)}\f^{(4)}_n + \left( \partial_i \dddot{\f}, \dddot{\f}, \dot{q} \dots \right) \,,
\end{align} 
where $\a^{i\a}_m$ is defined in (\ref{alpha}).
If one satisfies the primary conditions, one can get rid of the $\f^{(4)}_m$ terms and find
\begin{align}
E_{\f_m} + \frac{d}{dt}(V^\a_m E_{q_\a}) + \partial_i (\a^{i\a}_m E_{q_\a}) + U^\a_m E_{q_\a} = (S_i)_{(mn)}\partial_i \dddot{\f}_n + \left( \dddot{\f}, \dot{q} \dots \right) \,.
\end{align}
Hence, if the symmetric secondary conditions are satisfied, the mixed higher order terms $\partial_i\dddot{\f}_m$ also drop out leading to
\begin{align}
E_{\f_m} + \frac{d}{dt}(V^\a_m E_{q_\a}) + \partial_i (\a^{i\a}_m E_{q_\a}) + U^\a_m E_{q_\a} = S_{[mn]}\dddot{\f}_n +  \left( \ddot{\f}, \dot{q}, \dots \right) \,,
\end{align}
such that, if one satisfies the antisymmetric secondary conditions, one can lastly get rid of the $\dddot{\f}_m$ terms, getting equations containing at most $\ddot{\f}_m$ and $\dot{q}_\a$ (and up to second order spatial derivatives thereof).
Therefore it is again clear that one does not need to specify the naive amount of $4M + 2A$ initial conditions to solve the equations of motion, but rather only $2M + A$, thus leading to $M+A$ propagating degrees of freedom. 

\smallskip

Let us see how the presence of the additional, independent, symmetric secondary conditions modifies the dof counting (compared to the mechanics case) for partially degenerate theories where only the primary conditions are satisfied. If we have an even number of primary constraints there is no difference: there is an integer number of degrees of freedom. However, if we have an odd number of primary constraints, one generically has a {\it non-integer} number of degrees of freedom. This is due to the presence of the set of symmetric secondary conditions which, unlike the antisymmetric conditions, is not guaranteed to have a null eigenvalue. Therefore, generically no secondary constraints are present and a ``half'' degree of freedom propagates.
This pathology is known to be present in some Lorentz breaking modifications of GR, such as Horava--Lifschitz \cite{Henneaux:2010vx} or Lorentz breaking massive gravity \cite{Comelli:2012vz}.

\subsection{Lorentz invariant theories}
\label{LIsec}

\no So far we have made no assumptions concerning possible global symmetries the theories might have. In this section we consider the case of Lorentz invariant theories. We restrict ourselves to the case where all the fields are scalars under Lorentz transformations. Indeed, if one wants theories that solely propagate healthy spin 1 or 2 degrees of freedom, one is automatically led to additional degeneracies, already in the healthy sector. For example, describing a massless spin 1 degree of freedom via a vector field, necessarily implies the existence of a $U(1)$ gauge symmetry, which goes beyond our ansatz. Similarly a massless spin 2 degree of freedom implies diffeomorphism invariance, again going beyond our assumptions. Also the massive spin 1 / spin 2 degrees of freedom imply additional degeneracies in the healthy sector (although they are not of the gauge type). Therefore to stay in our setup, we restrict ourselves to Lorentz invariant scalar field theories.

\smallskip

Let us start looking at what Lorentz invariance implies in this case. By definition we get
 \begin{align}
 \bar{L}(\f,\partial_\m \phi, \partial_\m\partial_\n \f) &\equiv L(\f,(\L^{-1})_\m^{~\r} \partial_\r \phi,(\L^{-1})_\m^{~\r}(\L^{-1})_\n^{~\s} \partial_\r\partial_\s \f) \nb \\
 &= L(\f,\partial_\m \phi, \partial_\m\partial_\n \f) + \partial_\m (J^\m(\f, \partial_\m \f)) \,,
 \end{align}
 where in the first line the fields are evaluated at $x' = \L\, x$, and in the second line  at $x$.
 However, in the following the dependence on space-time of the various fields will be understood.
 %However, from here on we ignore this, and are not bothered at which space-time points the fields are evaluated. Rather, we are only interested in the functional dependence of the Lagrangian on the fields.
 Using an infinitesimal form $(\L^{-1})_\m^{~\n} = \delta_\m^\n - \o_\m^{~\n}$ and subsequently expanding the left and right hand side to first order, we find 
 \be
 \bar{L} = L + \delta L = L + \partial_\m \delta J^\m \,,
 \ee
 where
 \be
 \delta L = L_{\f_m}\delta \f_m + L_{\partial_\m \f_m}\delta \partial_\m \f_m + L_{\partial_\m\partial_\n \f_m}\delta \partial_\m\partial_\n \f_m + L_{q_\a}\delta q_\a + L_{\partial_\m q_\a}\delta \partial_\m q_\a \,,
 \ee
 and
 \begin{align}  
 \delta \f_m &= \delta q_\a =  0, \qquad \delta \partial_\m \f_m = -\omega_\m^{~\r} \partial_\r \f_m \,, \nb \\
 \delta \partial_\m q_\a &= -\omega_\m^{~\r} \partial_\r q_\a, \qquad \delta \partial_\m\partial_\n \f_m = -\omega_\m^{~\r} \partial_\r\partial_\n \f_m - \omega_\n^{~\r} \partial_\m\partial_\r \f_m \,.
 \end{align}
Now, since the theory is Lorentz invariant, a Lorentz transformation does not change the degeneracy structure and
\be
\bar{P}_{(mn)} = P_{(mn)} + \delta P_{(mn)} = 0 \,,
\ee
where
\begin{align} 
 \delta P_{(mn)} &= v^A_m (\delta L)_{\dot{\psi}_A \dot{\psi}_B}v^B_n \nb \\
 &=  (\delta L)_{\ddot{\f}_m \ddot{\f}_n} + (\delta L)_{\ddot{\f}_m \dot{q}_\a}V_n^\a + V_m^\b (\delta  L)_{\ddot{\f}_n \dot{q}_\b} + V_m^\a  (\delta L)_{\dot{q}_\a \dot{q}_\b}  V_n^\b \,.
 \end{align}
%$\bar{L}$ should satisfy its own primary conditions
%\begin{align} 
%\bar{P}_{(mn)} \equiv \bar{L}_{\ddot{\f}_m \ddot{\f}_n} + \bar{L}_{\ddot{\f}_m \dot{q}_\a}\bar{V}^\a_n = 0, \qquad \bar{V}^\a_m &\equiv  -\bar{L}_{\ddot{\f}_m \dot{q}_\b}\bar{L}^{-1}_{\dot{q}_\b \dot{q}_\a}
%\end{align} 
%Next consider the variation of the primary condition, i.e.
% \begin{align} 
%\delta (P_{(mn)}) &\equiv \bar{P}_{(mn)} -  P_{(mn)} \\
%&=  (\delta L)_{\ddot{\f}_m \ddot{\f}_n} + (\delta L)_{\ddot{\f}_m \dot{q}_\a}V_n^\a + V_m^\b (\delta  L)_{\ddot{\f}_n \dot{q}_\b} + V_m^\a  (\delta L)_{\dot{q}_\a \dot{q}_\b}  V_n^\b
%\end{align} 
%where we used that at first order
%\begin{align} 
%\bar{V}^\a_m &= V_m^\a - \left( (\delta  L)_{\ddot{\f}_m \dot{q}_\b} + V_m^\gamma (\delta L)_{\dot{q}_\gamma \dot{q}_\b} \right) L^{-1}_{\dot{q}_\b \dot{q}_\a} \,.
%\end{align} 
Therefore, if the primary conditions are satisfied, also $\delta P_{(mn)}$ vanishes.
Considering the boost transformation in the $i$-direction, and denoting the corresponding variation by $\delta_i$, it follows that
%\begin{align}
%0 & =(\delta_{i} L)_{\ddot{\f}_m\ddot{\f}_n} + V_{m}^{\a}(\delta_{i}  L)_{\dot{q}_\a\ddot{\f}_n} + (\delta_{i}  L)_{\ddot{\f}_m \dot{q}_\b}V_{n}^\b +   V_{m}^\a(\delta_{i}  L)_{\dot{q}_\a\dot{q}_\b}V_{n}^\b \nb \\
%&= (P_{(mn)})_{\partial_i\Psi_j}\dot{\Psi}_j + (P_{(mn)})_{\dot{\Psi}_j}\partial_i\Psi_j + (S_i)_{(mn)} \,,
%\end{align}
\be
0 = \delta_i P_{(mn)} = (P_{(mn)})_{\dot{\Psi}_j}\partial_i\Psi_j + (P_{(mn)})_{\partial_i\Psi_j}\dot{\Psi}_j  + (S_i)_{(mn)} \,,
\ee
where we introduced the notation $\Psi \equiv \{\phi_m,\partial_\m \phi_m,\partial_\m \phi_m, q_\a\}$.
Hence if the primary conditions are satisfied, automatically the symmetric secondary conditions are satisfied as well. Therefore, in Lorentz invariant theories, only the primary and antisymmetric secondary conditions remain, much resembling the mechanics case.

\smallskip

At the level of the equations of motion, this means that if one can get rid of the fourth order time derivative terms  $\f^{(4)}_m$ in $E_{\f_m}$, then one can automatically also get rid of the mixed terms $\partial_i \dddot{\f_m}$.  Let us note however that, in general, this cannot be done in a Lorentz covariant manner. This is because the combinations
\begin{align} 
E_{\f_m} + \frac{d}{dt}(V^\a_m E_{q_\a}) + \partial_i (\a^{i\a}_m E_{q_\a}) + U^\a_m E_{q_\a} \,,
\end{align} 
are Lorentz invariant only if $W_m^{\m\a} \equiv (V_m^\a,\a^{i\a}_m)$ is a Lorentz vector and $U_m^\a$ is  a Lorentz scalar, which, in general, is not the case. An example of such a theory is given in the next section (see eq.~(\ref{Wnotvector})). Therefore, there is generically a tradeoff between manifest Lorentz invariance (LI) and manifestly lower order equations of motion: either the equations are manifestly Lorentz invariant and higher order, or the equations are not manifestly Lorentz invariant but lower order. Of course, there are also theories for which it \textit{can} be done in a Lorentz covariant manner.  This different behavior divides the set of healthy LI higher derivative theories in two subclasses. We will come back to this point in the next section.

\smallskip

Let us conclude by highlighting an important property of the number of degrees of freedom for partially degenerate Lorentz invariant theories. As noted, the structure of the constraint conditions for Lorentz invariant theories much resembles the one of mechanical systems. Since the symmetric secondary conditions are automatically satisfied if the primary conditions are, the counting of dof goes in the same way as for the mechanics case: one always has an integer number of degrees of freedom.
We have thus explicitly shown how Lorentz invariance protects from the propagation of ``half'' dof. This is relevant for many theories of interest where there is a single (second class) primary constraint.
In these theories, one does not need to check the existence of a companion secondary constraint in order to completely remove the ghost, as its presence is assured as a consequence of Lorentz invariance.
We expect that this property still holds for more general cases that go beyond the present analysis of scalar theories;  examples of this kind are dRGT massive gravity \cite{deRham:2010kj} and degenerate scalar-tensor theories \cite{BenAchour:2016fzp}.

\section{Analysis of degeneracy classes}
\label{classification}

\no Having derived the conditions needed to ensure the absence of ghosts in higher derivative theories, we will provide a formal classification according to generic structures one finds within the class of healthy higher derivative theories\footnote{Due to the very complicated nature of the conditions (they constitute a set of highly nonlinear coupled partial differential equations), they cannot be solved in full generality. One could restrict oneself to theories polynomial in $\ddot{\f}_m$ and $\dot{q}_\a$, and do an order by order analysis in the number of fields and the power of the derivative terms.  However, this quickly becomes intractable due to the large amount of functional freedom in the general and LI case, again leading to many conditions on these functions given as sets of coupled differential equations that cannot be easily solved. We therefore refrain from such an analysis.}. In particular we will argue that one should distinguish the following dependences of the nullvectors (\ref{eigenvectors}):
\begin{itemize}
\item \textbf{Class I}: $V_{m}^{\a} = 0$.
\item \textbf{Class II}: $V_{m}^{\a} = V_{m}^{\a}(\phi_n,\partial_\m\phi_n,q_\b)$.
\item \textbf{Class III}: $V_{m}^{\a} =V_{m}^{\a}(\phi_n,\partial_\m\phi_n,q_\b,\partial_\m\partial_\n \f_n, \partial_\m q_\b)$.
\end{itemize}
Note that we defined the classes to be disjoint. For each class we will focus on the structure of the constraints and address the question under what conditions field redefinitions and/or (extended) contact transformations\footnote{An extended discussion about our terminology and the possible redefinitions can be found in Appendix \ref{redefinitions}.} can put the theories in standard or simpler forms. Again we will consider mechanical systems, generic Lorentz violating field theories and Lorentz invariant field theories.

\subsection{Trivial constraints (Class I)}
\label{classI}

\no If $V_m^\a$ vanishes, there is no coupling between $\ddot{\f}_m$ and $\dot{q}_\a$, and hence the degeneracy is fully contained in the higher derivative sector and not due to the coupling to a healthy sector. In the Hamiltonian picture, the constraints are simply given by the conjugate momenta of the higher order fields. Since the primary conditions reduce to $L_{\ddot{\f}_m\ddot{\f}_n}=0$, these theories are necessarily linear in second order time derivatives. In fact, from the simplified secondary conditions, one can see that the equations of motion are automatically free of problematic terms, i.e. they contain at most second order time derivatives of the fields (although they can contain mixed higher order terms like $\partial_i \ddot{\f}_m$, etc.).

\smallskip

In the case of mechanical systems this class is particularly simple. The primary conditions imply linearity in $\ddot{\f}_m$,
\begin{align} 
L_{I}(\ddot{\f}_m,\dot{\f}_m,\f_m,\dot{q}_\a,q_\a) = \ddot{\f}_n f^n(\dot{\f}_m,\f_m, q_\a) + g(\dot{\f}_m,\f_m,\dot{q}_\a,q_\a) \,,
\end{align}
whereas the secondary conditions, $f^m_{\dot{\f}_n} = f^n_{\dot{\f}_m}$, ensure the existence of a function, $F(\dot{\f}_m,\f_m,q_\a)$, such that $F_{\dot{\f}_m} = f^m$. As a result the terms linear in $\ddot{\f}_m$ can be absorbed in a total derivative and one concludes that Class I is actually equal to the class of first order Lagrangians modulo total derivatives:
\begin{align} 
L_{I}(\ddot{\f}_m,\dot{\f}_m,\f_m,\dot{q}_\a,q_\a) = L(\dot{\f}_m,\f_m,\dot{q}_\a, q_\a) + \frac{d}{dt}F(\dot{\f}_m,\f_m,q_\a) \,,
\end{align}  
and as such no truly higher derivatives are present in this class.  

\smallskip

Turning to field theories, the primary conditions again imply linearity 
\begin{align}
L_I (\partial_\m \partial_\n \f_m,\partial_\m\f_m,\f_m,\partial_\m q_\a, q_\a) &= \ddot{\f}_n f^{n}(\partial_i\partial_\m\f_m,\partial_\m \f_m, \f_m,\partial_i q_\a, q_\a)  \\
&+ g(\partial_i \partial_\m \f_m,\partial_\m\f_m,\f_m,\partial_\m q_\a,q_\a) \,,
\end{align} 
and $f^m$ now has to satisfy the two secondary conditions
\begin{align} 
0 &= \frac{\partial f^{m}}{\partial(\partial_i \dot{\f}_n)} +  \frac{\partial f^{n}}{\partial(\partial_i \dot{\f}_m)} \,, \label{secsymCI} \\
0&=\frac{\partial f^{m}}{\partial \dot{\f}_n} - \frac{\partial f^{n}}{\partial \dot{\f}_m} - \frac12 \partial_i \left( \frac{\partial f^{m}}{\partial(\partial_i \dot{\f}_n)} - \frac{\partial f^{n}}{\partial(\partial_i \dot{\f}_m)} \right) \,. \label{secasymCI}
\end{align} 
It is not clear whether one can always find a total derivative that removes the $\ddot{\f}_m$ terms, as in the case of mechanical systems. Indeed a suitable total derivative should be of the form
\begin{align} 
\frac{d}{dt}F(\partial_i\partial_\m \f_m,\partial_\m \f_m, \f_m) &= F_{\dot \phi_m}\ddot{\f}_m + F_{\de_i \dot \phi_m}\partial_i \ddot{\f}_m + \dots \\
&\approx  \left( F_{\dot \phi_m} - \de_i F_{\de_i \dot \phi_m} \right) \ddot{\f}_m + \dots
\end{align} 
and hence one must require that $\left( F_{\dot \phi_m} - \de_i F_{\de_i \dot \phi_m} \right) = f^{m}$. We do not know whether for any $f^m$ satisfying the secondary conditions (\ref{secsymCI}) and (\ref{secasymCI}), such a function $F$ exists. 

We note that if $f^{m}$ does not depend on $\partial_i\dot{\f}_n$ (which is always the case when only one higher derivative field is present), condition (\ref{secsymCI}) disappears and (\ref{secasymCI}) reduces to that of mechanics. As a consequence a total derivative, that puts the theory in a manifestly healthy form, can always be found.

\smallskip

Lastly, let us consider field theories that are manifestly Lorentz invariant. Since the equations of motion are also manifestly Lorentz invariant, they do not contain any higher order mixed terms, and are thus purely second order.\footnote{This implies that not only $V^\a_m=0$ but also $\a^{i\a}_m = 0$, since if $V^\a_m$ vanishes then $E_{q_\a} =  -\a^{i\b}_{ m}L_{\dot{q}_\b \dot{q}_\a}\partial_i\ddot{\f}_m + \left( \dots \right)$.} Therefore, this class corresponds to the most general set of Lorentz invariant scalar field theories that yield second order equations of motion, and thus contains multi-Galileons \cite{Deffayet:2010zh,Padilla:2010de,Hinterbichler:2010xn} and their known generalizations \cite{Padilla:2012dx,Allys:2016hfl}. At the present time it is unknown what the most general form of such theories is, however as shown in \cite{Sivanesan:2013tba}, they are polynomial in second derivatives and have a particular antisymmetric structure. This antisymmetric structure implies that $f^m$ never depends on $\partial_i\dot{\f}_m$ and thus Lorentz invariant theories can always be rewritten in a manifestly healthy form via a total derivative. Of course, this total derivative does not need to respect manifest Lorentz invariance.

\subsection{Linear constraints (Class II)}
\label{ClassII}

In this class, in contrast to the former one, there is a nontrivial coupling between the healthy and higher derivative sector. This nontrivial coupling is responsible for the higher order terms in the equations of motion, although, as we have seen in the previous section, one can always get rid of these terms. 
In the Hamiltonian picture the constraints are given by linear combinations of the conjugate momenta.
Naively one would expect that Class II truly goes beyond Class I, however it turns out that one can always perform a field redefinition to put a theory in Class II in a form belonging to Class I: one can always disentangle the higher derivative sector from the healthy one.

To be precise, we will show that $V_{m}^{\a} = V_{m}^{\a}(q_\b, \phi_n, \partial_\m\phi_n)$ if and only if there exists an invertible field redefinition of the form
\begin{align}
\bar{q}_\a = \bar{q}_\a(q_\b, \phi_n, \partial_\m\phi_n) \,,
\end{align}
such that 
\begin{align}
L_{II}(\partial_\m\partial_\n \f_m,\partial_\m\f_m,\f_m,\partial_\m q_\a,q_\a) = \bar{L}_I (\partial_\m\partial_\n \f_m,\partial_\m\f_m,\f_m,\partial_\m \bar{q}_\a,\bar{q}_\a) \,.
\end{align} 

Necessity is easily established by starting from a theory in Class I, performing such a field redefinition and observing that $V^\a_{m} = -\frac{\partial\bar{q}_\b}{\partial\dot{\f}_m}(\frac{\partial\bar{q}_\b}{\partial q_\a})^{-1} $, and thus $V_{m}^{\a} = V_{m}^{\a}(q_\b, \phi_n, \partial_\m\phi_n)$.

Sufficiency requires a bit more work. Consider the following system of partial differential equations
\begin{align}
\frac{\partial u}{\partial\dot{\f}_m} + V_{m}^{\b}(q_\a,\f_n,\partial_\m\f_n)\frac{\partial u}{\partial q_\b} =0 \,.
\label{pdesystem}
\end{align} 
Applying Frobenius' theorem one finds that it has $A$ independent solutions, call them $\bar{q}_\a$, if and only if the following integrability conditions are satisfied
\begin{align} 
0 = \frac{\partial V_{n}^{\b}}{\partial\dot{\f}_m} - \frac{\partial V_{m}^{\b}}{\partial\dot{\f}_n} + V_{m}^{\a}\frac{\partial V_{n}^{\b}}{\partial q_\a} - V_{n}^{\a}\frac{\partial V_{m}^{\b}}{\partial q_\a} \equiv {\cal F}_{mn}^{\b} \,. \label{Frobenius}
\end{align}
Explicitly calculating these conditions, using the specific dependence of $V_{m}^{\a}$ and the fact that $L_{II}$ satisfies the primary conditions, we obtain
\begin{align} 
{\cal F}_{mn}^{\b} = L^{-1}_{\dot{q}_\b \dot{q}_\a} \frac{\partial}{\partial\dot{q}_\a}S_{[mn]} \,.
\end{align} 
Therefore it vanishes by virtue of the antisymmetric secondary conditions. By subsequently using the nondegeneracy of the healthy sector and the fact that $\bar{q}_\a$ are independent, one can conclude that $\frac{\partial \bar{q}_\a}{\partial q_\b}$ is invertible.
Thus there always exists an invertible field redefinition $\bar{q}_\a$ that satisfies (\ref{pdesystem}). Now let $\bar{L}(\partial_\m\partial_\n \f_m,\partial_\m\f_m,\f_m,\partial_\m \bar{q}_\a,\bar{q}_\a) \equiv L_{II}(\partial_\m\partial_\n \f_m,\partial_\m\f_m,\f_m,\partial_\m q_\a,q_\a)$, then their null vectors are related as
\begin{align}
\bar{V}_{m}^{\a} &= \frac{\partial \bar{q}_\a}{\partial\dot{\f}_m} + V_{m}^{\b}\frac{\partial \bar{q}_\a}{\partial q_\b} \,.
\end{align}
Thus, since $\bar{q}_\a$ satisfies (\ref{pdesystem}), we observe that $\bar{V}_{m}^{\a} = 0$ and the Lagrangian $\bar{L}$ belongs to Class~I, concluding our proof.  

\smallskip

Turning to manifestly Lorentz invariant theories we note that, although they can be mapped to Class I via the above field redefinition, this transformation does not need to be compatible with manifest Lorentz invariance. That is, the transformed Lagrangian might not be manifestly Lorentz invariant. As we show in Appendix \ref{LIfieldredef}, a Lorentz invariant field redefinition exists if and only if $W_m^{\m \a} \equiv (V_{m}^{\a}, \alpha^{i\a}_{m})$ is a Lorentz vector and 
\begin{align} 
\frac{\partial W^{\m\b}_n}{\partial \partial_\n{\f}_m} -  \frac{\partial W^{\n\b}_m}{\partial \partial_\m{\f}_n} +  W^{\n\a}_m \frac{\partial W^{\m\b}_n}{\partial q_\a} - W^{\m\a}_n \frac{\partial W^{\n\b}_m}{\partial q_\a} = 0 \,.
\end{align} 
Therefore, any theory for which this is the case, is related to the most general, generalized multi-Galileon theory via a Lorentz invariant field redefinition, and thus does not truly go beyond the second order equations of motion ansatz.
In the opposite case instead, they really go beyond these theories. To see that this set is non-empty, consider for example the following bi-scalar theory
\begin{align} 
L_{II} = (q \square \f + 2 \partial_\m q \partial^\m \f)^2 \,, \label{Wnotvector}
\end{align} 
for which one can easily check that it is healthy and $W^\m$ is not a Lorentz vector.
Analogous theories in the context of degenerate scalar-tensor theories are those that cannot be mapped to Horndeski Lagrangians through generalized conformal and disformal transformations \cite{Crisostomi:2016czh, Achour:2016rkg, BenAchour:2016fzp}.

\subsection{Nonlinear constraints (Class III)}
\label{ClassIII}

The dependence of the nullvectors on $\dot{q}$ and $\ddot{\f}$, implies that the constraints in the Hamiltonian picture are nonlinear, in contrast to the linear ones of Class II. This has several implications regarding the structure of these theories.

To examine things further let us focus on mechanical systems, and in particular those systems with only one higher derivative variable but $A$ healthy variables. In this case the primary conditions reduce to the homogeneous Monge-Ampere equation in A dimensions and a general solution (for which $V_{\a}$ depends on $\ddot{\f}$ and $\dot{q}$) can be given in parametric form~\cite{Fairlie:1994in, Comelli:2013txa}. The secondary conditions are then automatically satisfied as explained in Section \ref{Field theories}. This parametric form is given by
\be
L = \ddot{\phi}\, \ll + \E + \frac{\de \ll}{\de V_\a} \Q^\a \,. \label{MAlagform}
\ee
Here $\mathcal{L}$ and $\E$ are arbitrary functions of the nullvector $V_\a$ and also $\f$, $\dot\f$ and $q$, and 
\be
\Q^\a = - \left(\frac{\de^2 \ll}{\de V_\a \de V_\b}\right)^{-1} \frac{\de \E}{\de V_\b} \,;
\ee
in turn $V_\a$ has to satisfy the following relation
\be
\ddot{\phi} \,V_\a + \Q^\a (V_\b) = \dot{q}_\a \,. \label{newfields}
\ee
To obtain explicit solutions, one first chooses the functions $\ll$ and $\E$ and subsequently solves~(\ref{newfields}) for $V_\a (\ddot{\phi}, \dot q_\b,\dot{\f},\f,q_\b)$.  Then plugging it into (\ref{MAlagform}), one obtains an explicit Lagrangian in terms of the variables $\phi$ and $q_\a$.

Given this general solution, we will now examine whether one can put it into manifestly healthy forms via known transformations. Because it is easy to generate explicit examples we will focus on the $A=1$ case.
Let us first observe that, in contrast to Class~II, Class~III cannot be rewritten into a simpler class via the field redefinitions considered for Class~II. This can be seen by noting that the nullvectors of two theories (in any class) related via such transformations, $\bar{q} =\bar{q}(q,\f,\dot{\f})$, are related by
\begin{align} 
V = \left(\bar{V} -\frac{\partial\bar{q}}{\partial\dot{\f}}\right)\left(\frac{\partial\bar{q}}{\partial q}\right)^{-1} \,.
\end{align}  
Hence, starting from a theory in Class I/II, one always ends up in another theory in Class I/II. Therefore, starting from Class III, one always remains in Class III with these redefinitions. 

As shown in Appendix \ref{redefinitions}, there is a much larger set of transformations one can consider, namely (extended) contact transformations of the form
\begin{align}
\bar{t} &= a t + f(\f,\dot{\f},q), \\
\bar{\f} &=  g(\f,\dot{\f},q),\qquad \bar{\f}' = G(\f,\dot{\f},q), \qquad \bar{\f}'' = \mathcal{G}(\f,\dot{\f},\ddot{\f},q,\dot{q}),\\
\bar{q} &= h(\f,\dot{\f},q),\qquad \bar{q}' = H(\f,\dot{\f},\ddot{\f},q,\dot{q}),
\end{align}
where $f$ and $g$ must satisfy a set of differential equations given in equation (\ref{diffconditions}) and $G$, $\mathcal{G}$ and $H$ follow from $f$, $g$ and $h$. Starting from a theory in Class I, $\bar{L}_I$, and performing such a transformation (with $h_q,f_{\dot{\f}}\neq 0$), one obtains a theory in Class III, $L_{III}$. In particular one finds
\begin{align} 
L_{III}(\ddot{\f},\dot{\f},\f,\dot{q},q) = \frac{d\bar{t}}{dt} \bar{L}_I(\bar{\f}'',\bar{\f}',\bar{\f},\bar{q}',\bar{q}) \,,
\end{align}
whose nullvector is given by
\begin{align}
V = -\frac{\partial\bar{q}'}{\partial \ddot{\f}}\left(\frac{\partial\bar{q}'}{\partial \dot{q}}\right)^{-1} = \frac{C + \dot{q}}{D + \ddot{\f}}\,,   \label{Vtrans}
\end{align}
where 
\begin{align} 
\qquad C = \frac{(f_{\dot{\f}}h_\f - h_{\dot{\f}}f_\f)\dot{\f} - h_{\dot{\f}}a}{f_{\dot{\f}}h_q - h_{\dot{\f}}f_q} \,, \quad D = \frac{(h_q f_\f - f_q h_\f)\dot{\f} + h_q a }{f_{\dot{\f}}h_q - h_{\dot{\f}}f_q} \,.
\end{align} 
Generic choices for the function $\Q$ in (\ref{newfields}) however, yield nullvectors whose dependence on $\ddot{\f}$ and $\dot{q}$ is not of this form, and thus not every theory in Class III can be reached from Class~I. Interestingly, the simplest option, namely to select $\ll$ and $\E$ such that $\Q$ is linear in $V$, i.e. $\Q = B\, V - A$, yields
\begin{align} 
V = \frac{A(\f,\dot{\f},q) + \dot{q}}{B(\f,\dot{\f},q) +  \ddot{\f}} \,.
\end{align} 
However, it is not clear to us whether one can, for any $A$ and $B$, find a redefinition such that $C = A$ and $D =B$. Regardless, one concludes that at most a very small \textit{subset} of Class III can be mapped to Class I via these transformations.  

We expect that also for the general case of $M$ higher derivative variables and $A$ healthy variables, one cannot reduce all Class III theories to Class I. In fact, the effectiveness of contact transformations actually seems to be reduced in the case where more higher derivative fields are present, since no nontrivial contact transformations (i.e. those not following from point transformations) exist involving more than one dependent variable (i.e. multiple $\phi_m$). Although we have not analysed them in detail, we expect that the above results also apply to field theories, since they are generically more complicated. This also includes the specific subset of Lorentz invariant field theories, since they behave very much like mechanical systems. 

\smallskip

To summarise, most of the theories in Class III are intrinsically higher order and cannot be brought to a standard, first order form, via known transformations. This is due to the non-linear nature of their constraints.

\section{Conclusions}
\label{conclusions}

We have performed a constraints analysis of field theories with coupled degrees of freedom. Restricting to theories without gauge symmetries, we have derived the conditions in order to evade the Ostrogradsky ghosts. 
They amount to a set of symmetric {\it primary} conditions and two sets of {\it secondary} conditions, one symmetric and the other antisymmetric.
Remarkably, the symmetric secondary conditions are automatically enforced by Lorentz invariance,
explaining how it explicitly protects from the propagation of pathological ``half'' degrees of freedom.

\begin{figure}[ht]
\centering{
	\includegraphics[width=13cm]{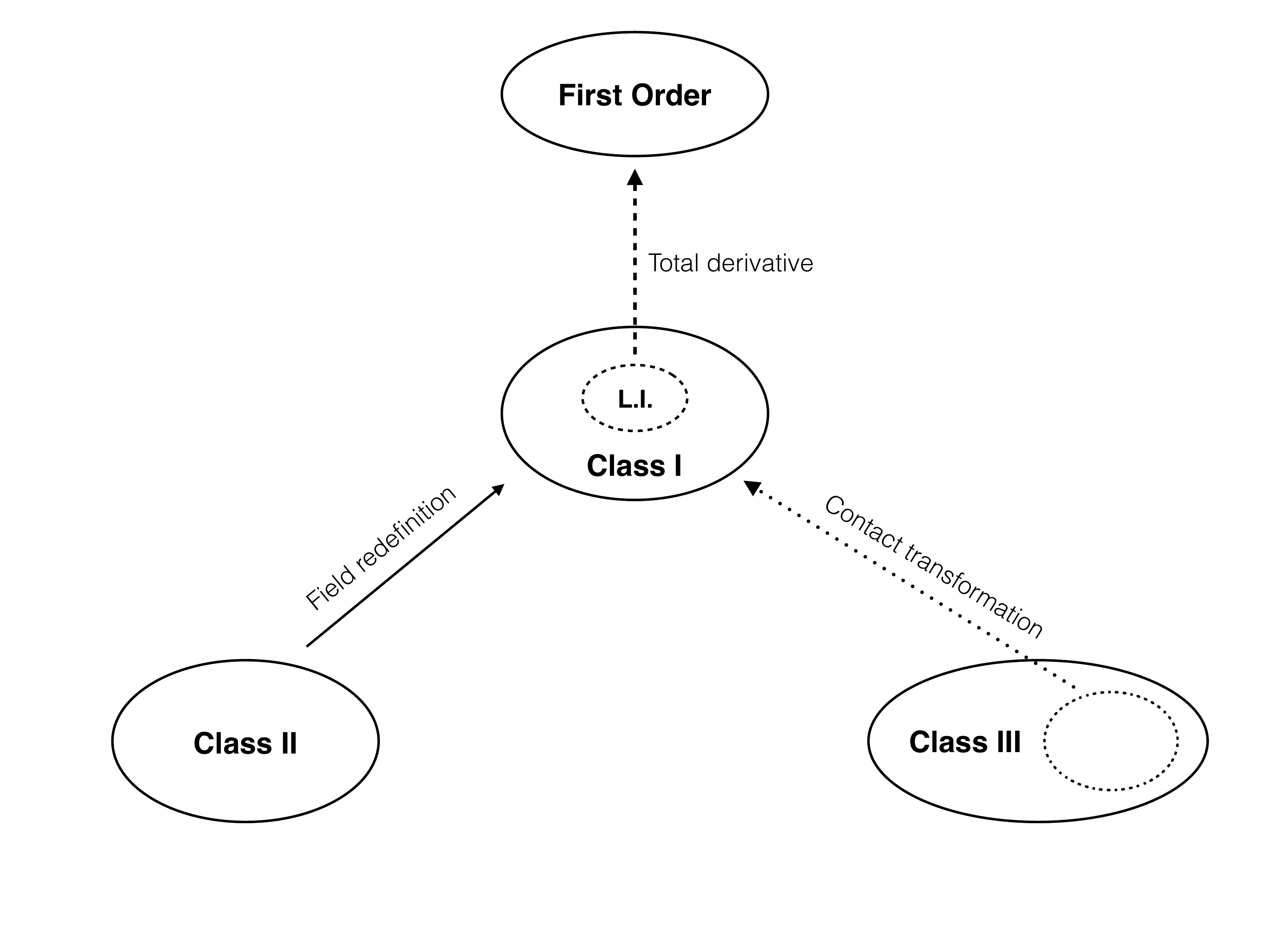}
}
\caption{\it Schematic representation of the three different classes of theories and their connections. Class II theories can always be put in Class I form via field redefinitions. Only a very small subset of theories in Class III can be brought to Class I with extended contact transformations. Finally, Lorentz invariant theories in Class I can be reduced to standard, first order form by adding a total derivative.}
\label{Classes}
\end{figure}

Secondly, we have outlined a number of classes of degenerate theories, depending on the properties of the null vector, and proved a number of equivalence relations between these classes. This classification is illustrated in Figure \ref{Classes} and its most salient features are:
 \begin{itemize}
 \item All Lorentz invariant field theories in Class I can be written in a manifestly healthy, first-order form, modulo a total derivative; however, one generically sacrifices manifest Lorentz invariance in doing so.
 \item All field theories in Class II can be brought to Class I by means of a field redefinition; again, this does not necessarily preserve manifest Lorentz invariance.
 \item Only a very small subset of theories in Class III can be brought to Class I by means of (extended) contact transformations.
\end{itemize}

%Our results can find concrete applications in many contexts. Beside model building, having provided the conditions for healthy multi-fields theories, 

We have thus illustrated the transformations that relate higher order theories to first order ones, and discussed their relation with manifest Lorentz invariance. Also we have shown which sub-classes of theories are instead intrinsically higher order and cannot be recast into a manifestly first order form by performing redefinitions and/or adding total derivatives. In particular this includes the majority of Class III.

For Lorentz invariant theories with a single higher derivative mode, we have shown how the required secondary constraint needed to completely remove the ghost, is always present when a primary constraint is present.
This applies in principle to beyond Horndesky theories, as well as dRGT massive gravity, saving one from a complicated analysis to confirm its existence.

Amongst the topics we have not touched there is the inclusion of degeneracies in the healthy sector, e.g.~arising from gauge symmetries or the absence of specific kinetic terms. This option would be necessary in order to go beyond scalar fields and discuss other (bosonic) Lorentz representations.  We expect the implications of Lorentz invariance regarding the structure of the constraints to be similar in such cases. Similarly, we have only included up to second order derivatives, whereas there are also healthy third or higher order theories. It is clear however, that the corresponding constraint analysis is significantly more involved than the one presently performed, and we leave this study for future work.

\section*{Acknowledgments}
MC is supported by the European Research Council through grant 646702 (CosTesGrav). RK acknowledges the Dutch funding agency 'Netherlands Organisation for Scientific Research' (NWO) for financial support.

\appendix 

\section{Lagrangian analysis}
\label{Laganalysis}
In this Appendix we perform the Lagrangian constraint analysis \cite{Algorithm1,Algorithm2} for the general Lagrangian (\ref{Lagstudied}):
\begin{align}
L(\partial_\m\partial_\n \f_m, \partial_\m \phi_m, \f_m,\partial_\m q_\a, q_\a) \,,
\end{align} 
and derive the conditions that, within our assumptions, are necessary and sufficient for the existence of the right amount of primary and secondary Lagrangian constraints to ensure that the Ostrogradsky degrees of freedom are eliminated. The analysis closely follows that of \cite{Klein:2016aiq}, where it has been done for mechanical systems. Let us give a short summary of the algorithm in the case of mechanical systems:
\begin{itemize}
\item First one puts the theory in a manifestly first order form by introducing suitable auxiliary fields. Then one calculates the equations of motion. These contain terms with second order time derivatives. One then determines whether combinations of the equations that do not contain second order time derivatives exist. These are the constraint equations. After having identified them, one evolves these constraint equations in time and adjoins these time derivatives with the original set of equations of motion, obtaining the 'equations of motion' which form the starting point of the next step in the algorithm.
\item  Next one repeats the analysis only for the larger set of 'equations of motion': one identifies possible additional constraint equations and subsequently evolves them in time to obtain the set of 'equations of motion for the next step.
\item This process is repeated until no further constraints are found. At this point the algorithm terminates and one has uncovered all constraints present in the theory.
\end{itemize}

Now, if one is considering field theories the algorithm is in essence the same, but one has to take into account the following points: 
\begin{itemize}
\item During any given step of the algorithm, spatial derivatives (of any order) of the 'equations of motion' of that given step are also allowed in forming possible new constraint equations. 
\item At any step of the algorithm the 'equations of motion' might contain, in addition to purely second order time derivatives, problematic terms involving spatial derivatives of second order time derivatives. Any constraint equation must of course be free of both types of problematic terms. As we will see the spatial derivatives of the 'equations of motion' play a key role in being able to achieve this. 
\end{itemize}

The degrees of freedom in the theory can be determined \cite{DOFcount1,DOFcount2,DOFcount3,DOFcount4} via $\#$ d.o.f. $= N - \frac{1}{2}l$ , where $N$ is the total number of fields (in a first order formulation) and  $l$ is the total number of constraint equations. Here one is assuming that no gauge symmetries are present in the theory.

\subsection{Non-degenerate Lagrangians}
\no First we put the theory in a first order form. This can be done in several equivalent ways, and we opt for the following:
\begin{align}
L(\partial_\m\partial_\n \f_m, \partial_\m \phi_m, \f_m,\partial_\m q_\a, q_\a) &\approx L(\dot{A}_m,\partial_i A_m,A_m, \partial_i\partial_j \f_m,\partial_i \f_m, \f_m,\dot{q}_\a,\partial_i q_\a) \nb \\
&+ \l^m (\dot{\f}_m - A_m) \,.
\end{align}
Now we can proceed with the constraint algorithm, starting off with determining the equations of motion
\begin{align}
E_{A_m} &\equiv L_{\dot{A}_m\dot{A}_n} \ddot{A}_n + L_{\dot{A}_m\dot{q}_\b}\ddot{q}_\b + L_{\dot{A}_m\psi}\dot{\psi} + L_{\partial_i A_m \chi}\partial_i\chi - L_{A_m} + \l^m \,, \\
E_{q_\a} &\equiv L_{\dot{q}_\a\dot{A}_n} \ddot{A}_n + L_{\dot{q}_\a\dot{q}_\b}\ddot{q}_\b + L_{\dot{q}_\a\psi}\dot{\psi} + L_{\partial_i q_\a \chi}\partial_i\chi - L_{q_\a} \,, \\
E_{\f_m} &\equiv -\partial_i\partial_j L_{\partial_i\partial_j \f_m} +\partial_i L_{\partial_i\f_m} - L_{\f_m} + \dot{\l}^m  \,, \\
E_{\l^m} &\equiv -(\dot{\f}_m - A_m) \,.
\end{align}
Here we introduced the short hand notation: $\chi \equiv \{\dot{A}_m,\partial_i A_m,A_m, \partial_i\partial_j \f_m,\partial_i \f_m, \f_m\,\dot{q}_\a,\partial_i q_\a \}$ and $\psi \equiv \chi \backslash\{\dot{A}_m,\dot{q}_\a\}$. If the Lagrangian is non-degenerate the only constraint equations are
\begin{align} 
C_{\f_m} \equiv E_{\f_m}  \,, \\
C_{\l^m} \equiv E_{\l^m} \,.
\end{align}
Time evolving them yields
\begin{align} 
\frac{d}{dt}C_{\f_m}&= \ddot{\l}^m + (C_{\f_m})_{\dot{A}_m} \ddot{A}_m + (C_{\f_m})_{\dot{q}_\a} \ddot{q}_\a  + ...\,, \\
\frac{d}{dt}C_{\l^m}&= -\ddot{\f}_m + ... \,.
\end{align}
Here we only included the terms that contain purely second order time derivatives, because it is already clear from these (specifically the $\ddot{\l}^m$ term) that no secondary constraint equations can be formed. Therefore the algorithm terminates and one concludes that in total $2M$ constraints are present, which are purely due to the redundant first order description. The theory thus propagates $3M + A - \frac{1}{2}(2M) = 2M + A$ degrees of freedom (of which $M$ are ghosts) as a non-degenerate higher derivative theory should.

\subsection{Degenerate Lagrangians}
\no Turning to the degenerate case, we see that in order to have $M$ additional primary constraints we must demand that 
\be
L_{\dot{A}_m\dot{A}_n} - L_{\dot{A}_m\dot{q}_\a}L^{-1}_{\dot{q}_\a\dot{q}_\b} L_{\dot{q}_\b \dot{A}_n} = 0 \,,\label{degcond}
\ee
which is equivalent to the existence of $M$ null vectors, $v_m^A = (\delta_m^n,V^\a_m)$, of the Hessian of $L$ w.r.t. $\dot{A}_m$ and $\dot{q}_\a$. Specifically we have
\begin{align} 
V_m^\a = -L_{\dot{A}_m\dot{q}_\b}L^{-1}_{\dot{q}_\b\dot{q}_\a} \,.
\end{align}
In terms of the original variables only, i.e. using the identification $A_m = \dot{\f}_m$, (\ref{degcond}) reduces to the primary conditions (\ref{primarycond}). The $M$ additional primary constraints are then given by:
\begin{align}
C_m &\equiv E_{A_m} + V_m^\a  E_{q_\a} \nb \\
&= (L_{\dot{A}_m\psi}+ V_m^\a L_{\dot{q}_\a\psi}) \dot{\psi} + (\partial_i L_{\partial_i A_m} +  V_m^\a \partial_i L_{\partial_i q_\a }) - (L_{A_m} + V_m^\a L_{q_\a}) \,.
\end{align}
Time evolving them yields
\begin{align}
\frac{dC_m}{dt} &= (C_m)_{\dot{A}_n} \ddot{A}_n + (C_m)_{\partial_i\dot{A}_n}  \partial_i \ddot{A}_n + (C_m)_{\dot{q}_\b} \ddot{q}_\b + (C_m)_{\partial_i\dot{q}_\b} \partial_i\ddot{q}_\b \nb \\
&+ (C_m)_{\dot{\f}_n} \ddot{\f}_n + (C_m)_{\partial_i\dot{\f}_n} \partial_i \ddot{\f}_n  + (C_m)_{\partial_i\partial_j\dot{\f}_n} \partial_i\partial_j \ddot{\f}_n + ... \,.
\end{align}
Next we must demand that $M$ secondary constraints exist in order to fully remove the ghost degrees of freedom.  The most general such constraints will have the following form:
\begin{align}
D_m &=  \frac{d}{dt}C_m + U^{\a}_m E_{q_\a} + \a^{i\a}_{m}\partial_i E_{q_\a} \nb \\
&+ (C_m)_{\dot{\f}_n}\frac{d}{dt}C_{\l^n} + (C_m)_{\partial_i\dot{\f}_n} \partial_i\frac{d}{dt}C_{\l^n} + (C_m)_{\partial_i\partial_j\dot{\f}_n} \partial_i\partial_j\frac{d}{dt}C_{\l^n} \,.
\end{align}
One can see this by first noting that no terms involving $E_{A_m}$ or its spatial derivatives are present since, by virtue of the primary conditions, their relevant higher order derivative terms are not independent of those of $E_{q_\a}$ and its spatial derivatives. In addition, no higher order spatial derivatives of the equations of motion are present, as these will actually introduce even higher order problematic terms.

Now, depicting the relevant higher order terms in these combinations yields:
\begin{align}
D_m &= \{(C_m)_{\dot{A}_n} + U^\a_m L_{\dot{q}_\a \dot{A}_n} + \a^{i\a}_m \partial_i L_{\dot{q}_\a \dot{A}_n}\} \ddot{A}_n + \{(C_m)_{\dot{q}_\b} + U^\a_m L_{\dot{q}_\a \dot{q}_\b} + \a^{i\a}_m \partial_i L_{\dot{q}_\a \dot{q}_\b}\}\ddot{q}_\b \nb \\
&+ \{ (C_m)_{\partial_i\dot{A}_n} + \a^{i\a}_m L_{\dot{q}_\a\dot{A}_n} \}\partial_i \ddot{A}_m + \{ (C_m)_{\partial_i\dot{q}_\b} + \a^{i\a}_m L_{\dot{q}_\a\dot{q}_\b}\}\partial_i \ddot{q}_\a + ... \,.
\end{align}
From this one can see that $U^\a_m$ and $\a^{i\a}_m$ exist such that all these terms vanish, if and only if the following conditions are met
\begin{align}
(C_m)_{\dot{A}_n} + (C_m)_{\dot{q}_\a}V^\a_n - (C_m)_{\partial_i\dot{q}_\b}L^{-1}_{\dot{q}_\b\dot{q}_\a} (\partial_i L_{\dot{q}_\a\dot{A}_n} + \partial_i L_{\dot{q}_\a\dot{q}_\r} V^\r_n) = 0 \,, \\
(C_m)_{\partial_i\dot{A}_n}+ (C_m)_{\partial_i\dot{q}_\b}V^\b_n = 0 \,.
\end{align}
Using explicit expressions we obtain
\begin{align}
0 &= (\partial_i L_{\partial_i A_m \dot{A}_n} + V_m^\a \partial_i L_{\partial_i q_\a \dot{A}_n} + \partial_i L_{\partial_i A_m \dot{q}_\b}V_n^\b + V_m^\a \partial_i L_{\partial_i q_\a \dot{q}_\b}V_n^\b) \nb \\
&+  \partial_i V^\b_n(L_{ \partial_i A_m\dot{q}_\b} + L_{\dot{A}_m \partial_i q_\b } + 2V_m^\a L_{\dot{q}_{(\a}\partial_i q_{\b)}}) \nb \\
&+ (L_{\dot{A}_m A_n} - L_{A_m \dot{A}_n}) + V_m^\a (L_{\dot{q}_\a A_n} - L_{q_\a\dot{A}_n}) \nb \\
& + (L_{\dot{A}_m q_\b} -L_{A_m\dot{q}_\b})V_n^\b  + V_m^\a (L_{\dot{q}_\a q_\b} -L_{q_\a\dot{q}_\b} )V_n^\b \label{second1} \,, \\
0 &= 2L_{\dot{A}_{(m}\partial_i A_{n)}} + 2V_{(m}^\a(L_{\dot{q}_\a \partial_i A_{n)}} + L_{ \partial_i q_\a \dot{A}_{n)}})  + 2V_m^\a L_{\dot{q}_{(\a}\partial_i q_{\b)}}V^\b_n \,, \label{second2}
\end{align}
and
\begin{align} 
U^\a_m &= ((C_m)_{\dot{q}_\b} - \a^{i\r}_m \partial_i L_{\dot{q}_\r \dot{q}_\b})L^{-1}_{\dot{q}_\b \dot{q}_\a} \,,\label{U} \\
\a^{i\a}_m &= - (L_{ \partial_i A_m\dot{q}_\b} + L_{\dot{A}_m \partial_i q_\b } + 2V_m^\r L_{\dot{q}_{(\r}\partial_i q_{\b)}})L^{-1}_{\dot{q}_\b\dot{q}_\a} \label{alpha} \,. 
\end{align} 
Therefore we conclude that if and only if the primary conditions (\ref{degcond}) hold, $M$ additional ($3M$ in total) primary constraint equations are present. Moreover, if and only if in addition the secondary conditions (\ref{second1}) and (\ref{second2}) are satisfied, $M$ secondary constraint equations exist. Assuming that no further conditions are imposed, no tertiary constraint equations will be present and the theory then propagates $3M + A - \frac{1}{2}(3M + M) = M + A$ degrees of freedom and the $M$ Ostrogradsky ghosts are not present.

Note that the symmetric part of (\ref{second1}) is in fact the spatial derivative of (\ref{second2}). Hence one ends up with one symmetric and one antisymmetric set of conditions, which, when written in terms of the original variables, precisely yield the symmetric (\ref{symmetricsecondary}) and antisymmetric (\ref{secondarycond}) secondary conditions.

\section{Hamiltonian analysis}
\label{Hamanalysis}

\subsection{Non-degenerate Lagrangians}

\no In this Appendix we perform the canonical analysis, using the Dirac method for constrained systems \cite{Dirac}, of the general Lagrangian (\ref{Lagstudied})
\be
L(\phi_m,\de_\mu \phi_m, \de_\mu \de_\nu \phi_m, q_\a, \de_\m q_\a) \approx L(\phi_m, A_\mu^m, \de_\nu A_\mu^m, q_\a, \de_\m q_\a) + \l_m^\mu (\de_\mu \phi_m - A_\mu^m) \label{Lag} \,.
\ee
Using the relations imposed by the Lagrangian multipliers $\l_m^\mu$, we have that $\de_\mu A_\nu^m = \de_\nu A_\mu^m $ and we can replace $\dot A_i^m = \de_i A_0^m$. To be precise, these relations hold only on-shell, i.e. on the phase space of constraints, however since they are second class constraints, they can be consistently imposed during the analysis.

Separating the space and time components, the Lagrangian (\ref{Lag}) becomes
\be
L = L(\phi_m, A_0^m, A_i^m, \dot A_0^m ,\de_i A_0^m, \de_i A_j^m, q_\a, \dot q_\a, \de_i q_\a) + \l_m^0 (\dot \phi_m - A_0^m) + \l_m^i (\de_i \phi_m - A_i^m) \,. \label{Lagfol}
\ee
The momenta conjugated to the fields and the primary constraints associated to the Lagrangian (\ref{Lagfol}) are

\begin{itemize}

\item $\pi_m \equiv \frac{\de L}{\de \dot \phi_m} = \l^0_m \qquad \Rightarrow \qquad (\pi_m - \l^0_m) \approx 0 \qquad M$ primary constraints  

\item $\L^0_m \equiv \frac{\de L}{\de \dot \l^0_m}  = 0 \,\,\,\qquad \Rightarrow \qquad\, \L^0_m \approx 0 \qquad\qquad\quad M$ primary constraints

\item $\L^i_m \equiv \frac{\de L}{\de \dot \l^i_m}  = 0 \,\,\,\qquad \Rightarrow \qquad\, \L^i_m \approx 0 \qquad\qquad\quad M \cdot i$ primary constraints

\item $P_i^m \equiv \frac{\de L}{\de \dot A_i^m}  = 0 \,\,\qquad \Rightarrow \qquad P_i^m \approx 0 \qquad\qquad\quad M \cdot i$ primary constraints

\item $P_0^m \equiv \frac{\de L}{\de \dot A_0^m} \,\qquad\qquad \Rightarrow \qquad \dot A_0^m = f^m(P_0^n, \phi_n, A_0^n, A_i^n,\de_i A_0^n, \de_i A_j^n, q_\a, \de_i q_\a, p_\a)$

\item $p_\a \equiv \frac{\de L}{\de \dot q_\a} \quad\qquad\qquad \Rightarrow \qquad\, \dot q_\a = g_\a(P_0^n, \phi_n, A_0^n, A_i^n,\de_i A_0^n, \de_i A_j^n, q_\b, \de_i q_\b, p_\b)$

\end{itemize}
{where $i$ refers to the number of spatial dimensions}. In the last two lines we have not assumed any extra degeneracy for the moment.
The total Hamiltonian is the sum of the canonical Hamiltonian plus the primary constraints enforced through multipliers
\be
H_T = H_C + \int d^3 x \left[ a_m(\pi_m - \l^0_m) + b_0^m \L_m^0 +  b_i^m \L_m^i + c_i^m P^m_i \right] \, ,
\ee
where $H_C = \int d^3 x \, {\cal H}_C$ and
\be
{\cal H}_C = P_0^m f^m + p_\a g_\a - L(\phi_n, A_0^n, A_i^n,\de_i A_0^n, \de_i A_j^n, q_\b, \de_i q_\b, f^n, g_\b) + \l^0_mA_0^m - \l^i_m (\de_i \phi_m - A_i^m) \,.
\ee
Here, $a_m, b_0^m, b_i^m, c_i^m$ are the multipliers used to enforce the primary constraints.

Evolving the primary constraints we get
\begin{itemize}
\item $\left\{\L_m^i, H_T \right\} = \de_i \phi_m - A_i^m \approx 0 \,\,\,\,\,\quad\qquad\qquad M \cdot i$ secondary constraints

\item $\left\{P_m^i, H_T \right\} = \frac{\de L}{\de A_i^m} - P_0^n \frac{\de f^n}{\de A_i^m} -\l^i_m \approx 0 \qquad M \cdot i$ secondary constraints

\item $\left\{\L_m^0, H_T \right\} = a_m - A_0^m \approx 0 \,\quad\qquad \Rightarrow \qquad a_m = A_0^m$

\item $\left\{\pi_m - \l^0_m, H_T \right\}  \approx 0 \qquad\qquad\qquad \Rightarrow \qquad b_0^m = \frac{\de L}{\de \phi_m} - P_0^n \frac{\de f^n}{\de \phi_m} -\de_i \l^i_m$

\end{itemize}
The evolution of $\L_m^i$ and $P_m^i$ gives $2\, M \cdot i$ secondary constraints, instead from the evolution of $\L_m^0$ and $(\pi_m - \l^0_m)$ we can solve for two (out of four) set of multipliers, namely $a_m$ and~$b_0^m$.

Finally we need to evolve the secondary constraints

\begin{itemize}

\item $\left\{\de_i \phi_m - A_i^m, H_T \right\}  \approx 0 \qquad\qquad\qquad \Rightarrow \qquad c_i^m = \left\{\de_i \phi_m, H_T \right\}$

\item $\left\{\frac{\de L}{\de A_i^m} - P_0^n \frac{\de f^n}{\de A_i^m} -\l^i_m, H_T \right\}  \approx 0 \qquad \Rightarrow \qquad b_i^m = \left\{\frac{\de L}{\de A_i^m} - P_0^n \frac{\de f^n}{\de A_i^m}, H_T \right\}$

\end{itemize}
All the multipliers are now completely determined and the procedure stops.
It is easy to verify that all these constraints are second class, indeed they are simply associated with the redundancy of description we have used to reduce the order of the Lagrangian.
We started with $2(3M + 2 M \cdot i + A)$ canonical variables and we found $2(M + M \cdot i)$ constraints, therefore we are left with $2(2 M + A)$ canonical dof, or $2 M + A$ physical dof.
As it is well known, $M$ of these dof are due to the higher derivative terms in the Lagrangian (\ref{Lag}) and usually are associated with instabilities.

The safest of the solutions is to require that none of them actually propagate, demanding the existence of $M$ extra primary constraints in the $(A^m_0, P^m_0)$ sector. Since we are not considering here gauge invariant theories, we will also need to demand that these primary constraints generate $M$ secondary ones.

\subsection{Degenerate Lagrangians}

\no As we have seen, the fields $A_i^m$ and $\l_m^i$ don't play any significant rule so can be ignored in the rest of the analysis.
Also, to simplify the notation, from now on we drop the suffix ``zero'' from $A_0$ and $P_0$.

Requiring the existence of extra $M$ primary constraints means that the system of momenta $P^m = \de L/\de \dot A^m$ cannot be inverted anymore and solved in terms of the velocities $\dot A^m$.
The constraints therefore take the form
\be
\chi^m \equiv P^m - F^m (A^n, \de_i A^n, q_\a, \de_i q_\a, p_\a) \approx 0 \,, \label{primary}
\ee
and need to be added to the total Hamiltonian as
\be
H_T = H_C + \int d^3x \, \xi_m \chi^m \,, 
\ee
where $\xi_m$ are the usual multipliers and we have omitted the other primary constraints already analysed in the former section as they do not interact with the new ones.

It can be shown \cite{Motohashi:2016ftl} that the existence of the constraints (\ref{primary}) is in one-to-one correspondence with the degeneracy of the Hessian matrix of the Lagrangian with respect to the velocities $\dot A^m$ and $\dot q_\a$, i.e. conditions~(\ref{degcond}).
Therefore, in order to have the primary constraints (\ref{primary}), our Lagrangian has to satisfy the conditions (\ref{degcond}).

The evolution of the constraints (\ref{primary}) gives
\be
\left\{\chi^m(x), H_T \right\}  = \left\{\chi^m(x), H_C \right\} + \left\{\chi^m(x), \int d^3 y \, \xi_n (y) \chi^n (y) \right\} \,, \label{primevo}
\ee
and the last term is composed by the following parts
\bea
\left\{P^m (x), \int d^3 y \, \xi_n (y) F^n (y) \right\}  &=& \left( - \frac{\de F^n}{\de A^m} + \de_i \frac{\de F^n}{\de (\de_i A^m)} \right) \xi_n + \frac{\de F^n}{\de (\de_i A^m)} \de_i \xi_n  \,,  \pagebreak[3] \label{PB1} \\[2ex]
\left\{F^m (x), \int d^3 y \, \xi_n (y) P^n (y) \right\}  &=&  \frac{\de F^m}{\de A^n} \xi_n + \frac{\de F^m}{\de (\de_i A^n)} \de_i \xi_n  \,, \pagebreak[3] \label{PB2} \\[2ex]
\left\{F^m (x), \int d^3 y \, \xi_n (y) F^n (y) \right\}  &=& \left( \frac{\de F^m}{\de q_\a} \frac{\de F^n}{\de p_\a} - \frac{\de F^m}{\de p_\a} \frac{\de F^n}{\de q_\a} \nb \right. \\
&+& \left. \frac{\de F^m}{\de (\de_i q_\a)} \de_i \frac{\de F^n}{\de p_\a} + \frac{\de F^m}{\de p_\a} \de_i \frac{\de F^n}{\de (\de_i q_\a)} \right) \xi_n \nb \\[1ex] 
&+& \left( \frac{\de F^m}{\de (\de_i q_\a)} \frac{\de F^n}{\de p_\a} + \frac{\de F^m}{\de p_\a} \frac{\de F^n}{\de (\de_i q_\a)} \right) \de_i \xi_n  \,. \label{PB3}
\eea

The Poisson brackets (\ref{primevo}) have therefore the form
\be
\left\{\chi^m(x), H_T \right\}  = \left\{\chi^m(x), H_C \right\} + S^{mn} \xi_n +  (S_i)^{mn} \de_i \xi_n \,,
\ee
and in order to give secondary constraints we need to remove their dependency from $\xi_n$.
This gives the new conditions $S^{mn} = (S_i)^{mn} = 0$, whose specific form is easily obtainable from equations (\ref{PB1}) -- (\ref{PB3}).  

Using the primary constraints (\ref{primary}), it is possible to relate the derivatives of $F^m$ to those of the Lagrangian, namely
\bea
\frac{\de F^m}{\de p_\a} &=& - V^m_\a \,, \qquad
\frac{\de F^m}{\de q_\a} = L_{\dot{A}_m q_\a} + L_{\dot{q}_\b q_\a} V^m_\b \,, \qquad 
\frac{\de F^m}{\de (\de_i q_\a)} = L_{\dot{A}_m \partial_i q_\a} + L_{\dot{q}_\b \partial_i q_\a} V^m_\b \,, \nb \\[2ex]
\frac{\de F^m}{\de A^n} &=& L_{\dot{A}_m {A}_n} + L_{\dot{q}_\a A_n} V^m_\a \,, \qquad
\frac{\de F^m}{\de (\de_i A^n)} = L_{\dot{A}_m \partial_i A_n} + L_{\dot{q}_\a \partial_i A_n} V^m_\a \,.
\eea
Finally, substituting these relations in the above conditions, we get exactly equations (\ref{second1}) and (\ref{second2}).

\section{Redefinitions}
\label{redefinitions}

In this Appendix we discuss the possible redefinitions (of fields as well as coordinates) that can relate seemingly different theories. 

Let us consider a Lagrangian, $\bar{L}(\bar{\f},\bar{\partial}_\m\bar{\f},\bar{\partial}_\m\bar{\partial}_\n\bar{\f},\bar{q},\bar{\partial}_\m\bar{q})$, where the fields are functions of barred space-time coordinates $\bar{x}^\m$. Now assume that the Lagrangian belongs to any of the three degeneracy classes as discussed in Section \ref{classification}. 
We would like to know whether theories belonging to one of the degeneracy classes can be mapped to standard and/or simpler forms, again belonging to one of the classes, via general local and invertible redefinitions of both the fields as well as the space-time coordinates. Such a general transformation is of the form
\begin{align}
\bar{x}^\m &= \bar{x}^\m[x^\n,\f_n,q_\b] \,, \nb\\
\bar{\f}_m(\bar{x}^{\m}) &= \bar{\f}[x^\n,\f_n,q_\b] \,, \nb\\
\bar{q}_\a(\bar{x}^\m) &= \bar{q}_\a[x^\n,\f_n,q_\b] \,, \label{fullygeneraltransformation}
\end{align}
where the brackets indicate functional dependence (so dependence on the derivatives of the fields is implicit). Performing such a redefinition, the Lagrangian transforms as
\begin{align} 
L[x^\m,\f_m,q_\a] = |\textup{det}\frac{\partial\bar{x}^\m}{\partial x^\n}| \bar{L}(\bar{\f},\bar{\partial}_\m\bar{\f},\bar{\partial}_\m\bar{\partial}_\n\bar{\f},\bar{q},\bar{\partial}_\m\bar{q}) \,.
\end{align}
In order for this transformed Lagrangian, $L$, to fall within the scope of our analysis, we must demand
\begin{align} 
L[x^\m,\f_m,q_\a] = L(\f_m,\de_\m\f_m,\de_\m\de_\n \f_m,q_\a,\de_\m q_\a) \,,
\end{align} 
and degeneracy of the Lagrangian then automatically follows from the invertibility of the performed transformation. Now, in order for this to be the case in general, i.e. modulo accidental cancellations, we must restrict ourselves to those transformations (\ref{fullygeneraltransformation}) for which $|\textup{det}\frac{\partial\bar{x}^\m}{\partial x^\n}|$, $\bar{\f}_m$, $\bar{\de}_\m \bar{\f}_m$, $\bar{\de}_\m\bar{\de}_\n \bar{\f}_m$, $\bar{q}_\a$, $\bar{\de}_\m \bar{q}_\a$ are all functions of $(\f_n,\de_\m\f_n,\de_\m\de_\n \f_n,q_\b,\de_\m q_\b)$.

We do not know what the most general such transformation is, but let us note some notable types of transformations that fall within this class. The first are of course the well known \textit{field redefinitions}, i.e. transformations that only mix the fields (and possibly their derivatives) amongst themselves, but do not allow for mixing with the space-time coordinates. As seen in Section \ref{ClassII}, these transformations are sufficient to analyse Class II.   

A less frequently considered type of transformations are the \textit{contact transformations}. An $n$-th order contact transformation is an invertible redefinition that maps a set of space-time coordinates, fields and derivatives $(x^\m, \psi_i, \partial \psi_i, ..., \partial^{n}\psi_i)$ to another set of new coordinates, fields and derivatives  $(\bar{x}^\m, \bar{\psi}_i, \bar{\partial} \bar{\psi}_i, ..., \bar{\partial}^{n}\psi_i)$. Here in principle any of the barred quantities, both the coordinates as well as the fields and derivatives, can depend on any of the unbarred quantities. The simplest such transformations are the $0$-th order contact transformations, i.e. the \textit{point transformations}, that only truly mix the space-time coordinates and the fields. Now, it turns out that in fact only very little nontrivial higher order contact transformations exist. It has been proven that all contact transformations involving more than one field, are prolongations of point transformations. In the case of a single field nontrivial 1st-order contact transformations \textit{do} exist\footnote{A notable example of such a contact transformation is Galileon duality \cite{deRham:2013hsa,deRham:2014lqa}, which allows one to relate different Galileon theories to each other.}, but all higher order ones are prolongations of 0th/1st-order transformations. As we show in Section~\ref{ClassIII}, extensions of these contact transformations play a role in the analysis of Class~III.

\subsection{Extended contact transformations in the $(\f(t),q(t))$ case}
Here we determine the most general transformation in the case of mechanical systems with a single higher derivative variable and a single healthy variable. Let us consider a Lagrangian, $\bar{L}(\bar{\f},\bar{\f}',\bar{\f}'',\bar{q},\bar{q}')$, belonging to any of the three degeneracy classes as discussed in Section \ref{classification}. Performing a general, invertible, redefinition
\begin{align}
\bar{t} &= \bar{t}[t,\f,q]  = \bar{t}(t,\f,\dot{\f},...,\f^{(n)},q,\dot{q},...,q^{(m)}) \,, \nb\\
\bar{\f}(\bar{t}) &= \bar{\f}[t,\f,q] = \bar{\f}(t,\f,\dot{\f},...,\f^{(p)},q,\dot{q},...,q^{(q)}) \,, \nb\\
\bar{q}(\bar{t}) &= \bar{q}[t,\f,q] = \bar{q}(t,\f,\dot{\f},...,\f^{(r)},q,\dot{q},...,q^{(s)}) \,,\label{generaltransformation}
\end{align}
the Lagrangian transforms as
\begin{align} 
L[t,\f,q] = \frac{d\bar{t}}{dt} \bar{L} \,.
\end{align}
As noted we should only consider redefinitions for which
\begin{align} 
L[t,\f,q] = L(\f,\dot{\f},\ddot{\f},q,\dot{q}) \,.
\end{align} 
In order for this to be the case in general, i.e. modulo accidental cancellations, we must demand that the same holds for $\frac{d\bar{t}}{dt}$, $\bar{\f}$, $\bar{\f}'$, $\bar{\f}''$, $\bar{q}$ and $\bar{q}'$. \\

Thus, first requiring that $\frac{d\bar{t}}{dt} = \frac{d\bar{t}}{dt}(\f,\dot{\f},\ddot{\f},q,\dot{q})$, yields
\begin{align}
\bar{t} = a t + f(\f,\dot{\f},q) \,,
\end{align}
where $f$ is arbitrary, and $a \neq 0$ is a constant. Next starting from
\begin{align} 
\bar{\f} &=  \bar{\f}(\f,\dot{\f},\ddot{\f},q,\dot{q}) \,,\\
\bar{q} &= \bar{q}(\f,\dot{\f},\ddot{\f},q,\dot{q}) \,,
\end{align} 
and demanding the same dependence for their first derivatives
\begin{align} 
\bar{\f}' = \frac{d\bar{\f}}{d\bar{t}} = (\frac{d\bar{t}}{dt})^{-1}(\bar{\f}_{\ddot{\f}}\dddot{\f} + \bar{\f}_{\dot{q}}\ddot{q} +... )  \,, \\
\bar{q}'= \frac{d\bar{q}}{d\bar{t}} =  (\frac{d\bar{t}}{dt})^{-1}(\bar{q}_{\ddot{\f}}\dddot{\f} + \bar{q}_{\dot{q}}\ddot{q}+ ...) \,,
\end{align} 
yields $\bar{\f}_{\ddot{\f}} = \bar{\f}_{\dot{q}} = \bar{q}_{\ddot{\f}} = \bar{q}_{\dot{q}} = 0$. Thus in fact we find
\begin{align} 
\bar{\f} &=  \bar{\f}(\f,\dot{\f},q) \,, \\
\bar{q} &= \bar{q}(\f,\dot{\f},q) \,.
\end{align} 
Subsequently calculating the second derivative of $\bar{\f}$ yields
\begin{align} 
\bar{\f}'' =  \frac{d^2\bar{\f}}{d\bar{t}^2} = (\frac{d\bar{t}}{dt})^{-1}(\bar{\f}'_{\ddot{\f}}\dddot{\f} + \bar{\f}'_{\dot{q}}\ddot{q} +... ) \,,
\end{align} 
from which we conclude that
\begin{align} 
0 &= \bar{\f}'_{\ddot{\f}} \qquad \Rightarrow \qquad 0 = (a + \bar{t}_\f \dot{\f} + \bar{t}_{\dot{\f}}\ddot{\f} + \bar{t}_q \dot{q})\bar{\f}_{\dot{\f}} -(\bar{\f}_\f \dot{\f} + \bar{\f}_{\dot{\f}}\ddot{\f} + \bar{\f}_q \dot{q})\bar{t}_{\dot{\f}} \,,  \\
0 &= \bar{\f}'_{\dot{q}}  \qquad  \Rightarrow \qquad 0 = (a + \bar{t}_\f \dot{\f} + \bar{t}_{\dot{\f}}\ddot{\f} + \bar{t}_q \dot{q})\bar{\f}_{q} -(\bar{\f}_\f \dot{\f} + \bar{\f}_{\dot{\f}}\ddot{\f} + \bar{\f}_q \dot{q})\bar{t}_{q} \,, 
\end{align} 
which can be rewritten as:
\begin{align} 
0 &=\bar{t}_q \bar{\f}_{\dot{\f}} -\bar{\f}_q \bar{t}_{\dot{\f}} \,, \nb \\ 
0 &= (a + \bar{t}_\f \dot{\f})\bar{\f}_{\dot{\f}}  - \bar{\f}_\f \dot{\f}\bar{t}_{\dot{\f}} \,, \nb \\
0& = (a + \bar{t}_\f \dot{\f})\bar{\f}_{q}  - \bar{\f}_\f \dot{\f}\bar{t}_{q} \,. \label{diffconditions}
\end{align} 
Thus we conclude that the only redefinitions that satisfy our demands are of the form
\begin{align}
\bar{t} &= a t + f(\f,\dot{\f},q) \,, \nb \\
\bar{\f} &=  g(\f,\dot{\f},q),\qquad \bar{\f}' = G(\f,\dot{\f},q), \qquad \bar{\f}'' = \mathcal{G}(\f,\dot{\f},\ddot{\f},q,\dot{q}) \,, \nb\\
\bar{q} &= h(\f,\dot{\f},q),\qquad \bar{q}' = H(\f,\dot{\f},\ddot{\f},q,\dot{q}) \,,
\end{align}
where $f$ and $g$ have to satisfy the differential equations (\ref{diffconditions}) and $G$, $\mathcal{G}$ and $H$ follow from $f$, $g$ and $h$. Of course one must also require invertibility of the transformation, which is precisely the case if one can solve  $\bar{\f}$, $\bar{\f}'$ and $\bar{q}$ for $\phi$, $\dot{\f}$ and $q$. Note that these transformations generally go beyond contact transformations since they do not map any set of $n$-th (and lower) order derivatives to a new set of $n$-th (and lower) order derivatives.

\subsection{Lorentz invariant field redefinitions} 
\label{LIfieldredef}

\no In this Appendix we prove the following statement: a manifestly Lorentz invariant theory $L_{II}(\partial_\m\partial_\n\f_m, \partial_\m\f_m,\f_m,\partial_\m q_\a, q_\a)$, belonging to Class II, can be put in a manifestly Lorentz invariant form $\bar{L}_I( \partial_\n\partial_\n\f_m, \partial_\m\f_m,\f_m,\partial_\m \bar{q}_\a, \bar{q}_\a)$ (with $\bar{q}_\a = \bar{q}_\a(q,\f,\de \f)$ being Lorentz scalars), if and only if $W^{\mu \a}_m \equiv (V^\a_m, \a^{i\a}_m)$ is a Lorentz vector and 
\begin{align} 
\frac{\partial W^{\m\b}_n}{\partial \partial_\n{\f}_m} -  \frac{\partial W^{\n\b}_m}{\partial \partial_\m{\f}_n} +  W^{\n\a}_m \frac{\partial W^{\m\b}_n}{\partial q_\a} - W^{\m\a}_n \frac{\partial W^{\n\b}_m}{\partial q_\a} = 0 \,. \label{consistencyW}
\end{align}

Let us start with necessity. Assume that both $L_{II}$ and $\bar{L}_I$ are manifestly Lorentz invariant and related via a field redefinition of the specified form. Since $\bar L_I$ is Lorentz invariant, not only $\bar{V}^\a_m = 0$ but also $\bar{\a}^{i\a}_m = 0$ (as noted in Section \ref{classI}). Then, by calculating $W^{\mu \a}_m$ one finds
\begin{align} 
\frac{\partial \bar{q}_\a}{\partial \partial_\m{\f}_m} +  W^{\m\b}_m \frac{\partial \bar{q}_\a}{\partial q_\b} = 0 \,. \label{Wdiffeq}
\end{align}
Therefore, since $\bar{q}_\a$ is Lorentz invariant, we conclude that $W^{\m \a}_m = -\left( \frac{\partial \bar{q}_\b}{\partial \partial_\m\f_m}\right)\left( \frac{\partial \bar{q}_\b}{\partial q_\a}\right)^{-1}$ is a Lorentz vector. Lastly, one notes that the consistency conditions corresponding to (\ref{Wdiffeq}) are precisely (\ref{consistencyW}), which are thus automatically satisfied. 

Now, for sufficiency we first note that since $W^{\m \b}_m$ is a Lorentz vector and $V^\b_m = V^\b_m(q_\a,\phi_n,\partial_\m\f_n)$, the most general form is given by
\begin{align} 
W^{\mu\b}_m (q_\a,\f_p,\partial_\m\f_p) = A^{n\b}_m \partial^\m \phi_n,\qquad A^{n\b}_m =  A^{n\b}_m(q_\a,\f_p,X_{p,q}), \qquad X_{p,q} \equiv \frac{1}{2}\partial_\m \f_p \partial^\m \f_q \,.
\end{align}
Plugging this specific expression into (\ref{consistencyW}) it follows that
\begin{align} 
\left(A_{n}^{m\b} - A_{m}^{n\b}\right)\eta^{\m\n} + \left(\frac{\partial A_{n}^{p\b}}{\partial X_{mq}} -\frac{\partial A_{m}^{q\b}}{\partial X_{np}} + A_m^{p\a} \frac{\partial A_{n}^{q\b}}{\partial q_\a} -A_n^{q\a} \frac{\partial A_{m}^{p\b}}{\partial q_\a} \right) \partial^\m{\f}_p \partial^\n{\f}_q  = 0 \,.
\end{align} 
Since both terms in parenthesis are Lorentz invariant, one sees that $A_{n}^{m\b} = A_{m}^{n\b}$. Next we observe that because the consistency conditions (\ref{consistencyW}) are satisfied, one can always find independent $\bar{q}_\a$ that satisfy (\ref{Wdiffeq}). Picking precisely such a redefinition and calculating its variation under Lorentz transformations yields
\begin{align} 
\delta \bar{q}_\a &= \frac{\partial \bar{q}_\a}{\partial \partial_\mu\f_m} \delta \partial_\m \f_m \nb \\
&= -W^{\m \b}_m \left(\delta \partial_\m \f_m\right) \frac{\partial \bar{q}_\b}{\partial q_\a} \nb \\
&= \left(A^{n\b}_m \partial^\m \phi_n \omega_{\m\n} \partial^\n \f_m\right)\frac{\partial \bar{q}_\b}{\partial q_\a} \nb \\
&= 0 \,,
\end{align} 
where we used the symmetry of $A_{n}^{m\b}$. Thus, we conclude that $\bar{q}_\a$ is a Lorentz scalar and hence a manifestly Lorentz invariant field redefinition (and so is its inverse). Starting from a manifestly Lorentz invariant theory and performing this redefinition one obtains a Lagrangian belonging to Class I (since (\ref{Wdiffeq}) implies that $\bar{W}^{\m\a}_m = 0$) that is also manifestly Lorentz invariant.


\begin{thebibliography}{99}



\bibitem{Ostrogradski} 
M. Ostrogradsky. Mem. Ac. St. Petersbourg VI 4 (1850) 385;
%\cite{Woodard:2015zca}
%\bibitem{Woodard:2015zca} 
  R.~P.~Woodard,
  %``The Theorem of Ostrogradsky,''
  arXiv:1506.02210 [hep-th].
  

  
  \bibitem{Algorithm1}
    E.~C.~G. Sudarshan and N.~Mukunda, 
    \textit{Classical Dynamics: A Modern Perspective}
    (John Wiley and Son, New York, 1974).
    
    \bibitem{Algorithm2}
    H.~J.~Rothe and K.~D.~Rothe,
    \textit{Classical and Quantum Dynamics of Constrained Hamiltonian Systems}
    (World Scientific Publishing, Singapore, 2010).
  

  
  \bibitem{Nicolis:2008in}
  A.~Nicolis, R.~Rattazzi and E.~Trincherini,
  %``The Galileon as a local modification of gravity,''
  Phys.\ Rev.\ D {\bf 79} (2009) 064036
  %doi:10.1103/PhysRevD.79.064036
  [arXiv:0811.2197 [hep-th]].
  
  \bibitem{Deffayet:2009mn} 
  C.~Deffayet, S.~Deser and G.~Esposito-Farese,
  %``Generalized Galileons: All scalar models whose curved background extensions maintain second-order field equations and stress-tensors,''
  Phys.\ Rev.\ D {\bf 80}, 064015 (2009)
  %doi:10.1103/PhysRevD.80.064015
  [arXiv:0906.1967 [gr-qc]].
  
\bibitem{Lovelock:1971yv} 
  D.~Lovelock,
  %``The Einstein tensor and its generalizations,''
  J.\ Math.\ Phys.\  {\bf 12}, 498 (1971).
  %doi:10.1063/1.1665613
  %%CITATION = doi:10.1063/1.1665613;%%
  %1190 citations counted in INSPIRE as of 12 Jan 2017
  

  
  \bibitem{deRham:2010kj} 
  C.~de Rham, G.~Gabadadze and A.~J.~Tolley,
  %``Resummation of Massive Gravity,''
  Phys.\ Rev.\ Lett.\  {\bf 106}, 231101 (2011)
  %doi:10.1103/PhysRevLett.106.231101
  [arXiv:1011.1232 [hep-th]].
  
  \bibitem{deRham:2011rn} 
  C.~de Rham, G.~Gabadadze and A.~J.~Tolley,
  %``Ghost free Massive Gravity in the St\'uckelberg language,''
  Phys.\ Lett.\ B {\bf 711}, 190 (2012)
  %doi:10.1016/j.physletb.2012.03.081
  [arXiv:1107.3820 [hep-th]].
  

  \bibitem{Heisenberg:2014rta} 
  L.~Heisenberg,
  %``Generalization of the Proca Action,''
  JCAP {\bf 1405}, 015 (2014)
  %doi:10.1088/1475-7516/2014/05/015
  [arXiv:1402.7026 [hep-th]].
  

  \bibitem{Horndeski:1974wa} 
  G.~W.~Horndeski,
  %``Second-order scalar-tensor field equations in a four-dimensional space,''
  Int.\ J.\ Theor.\ Phys.\  {\bf 10}, 363 (1974).
  
  
  %\cite{Kobayashi:2011nu}
  \bibitem{Kobayashi:2011nu}
  T.~Kobayashi, M.~Yamaguchi and J.~Yokoyama,
  %``Generalized G-inflation: Inflation with the most general second-order field equations,''
  Prog.\ Theor.\ Phys.\  {\bf 126} (2011) 511
  %doi:10.1143/PTP.126.511
  [arXiv:1105.5723 [hep-th]].
  %%CITATION = doi:10.1143/PTP.126.511;%%
  %357 citations counted in INSPIRE as of 27 Mar 2017
  
  
   %\cite{Deffayet:2009wt}
    \bibitem{Deffayet:2009wt}
    C.~Deffayet, G.~Esposito-Farese and A.~Vikman,
    %``Covariant Galileon,''
    Phys.\ Rev.\ D {\bf 79} (2009) 084003
    doi:10.1103/PhysRevD.79.084003
    [arXiv:0901.1314 [hep-th]].
    %%CITATION = doi:10.1103/PhysRevD.79.084003;%%
    %521 citations counted in INSPIRE as of 15 Feb 2017
  
  %\cite{Deffayet:2011gz}
  \bibitem{Deffayet:2011gz}
  C.~Deffayet, X.~Gao, D.~A.~Steer and G.~Zahariade,
  %``From k-essence to generalised Galileons,''
  Phys.\ Rev.\ D {\bf 84} (2011) 064039
 % doi:10.1103/PhysRevD.84.064039
  [arXiv:1103.3260 [hep-th]].
  %%CITATION = doi:10.1103/PhysRevD.84.064039;%%
  %413 citations counted in INSPIRE as of 27 Mar 2017
  
  
  

  
  \bibitem{Tasinato:2014eka} 
  G.~Tasinato,
  %``Cosmic Acceleration from Abelian Symmetry Breaking,''
  JHEP {\bf 1404}, 067 (2014)
  %doi:10.1007/JHEP04(2014)067
  [arXiv:1402.6450 [hep-th]].
  
  \bibitem{Hull:2015uwa} 
  M.~Hull, K.~Koyama and G.~Tasinato,
  %``Covariantized vector Galileons,''
  Phys.\ Rev.\ D {\bf 93}, no. 6, 064012 (2016)
  %doi:10.1103/PhysRevD.93.064012
  [arXiv:1510.07029 [hep-th]].
  
  %\cite{Chatzistavrakidis:2016dnj}
\bibitem{Thanasis}
  A.~Chatzistavrakidis, F.~S.~Khoo, D.~Roest and P.~Schupp,
  %``Tensor Galileons and Gravity,''
  arXiv:1612.05991 [hep-th].
  %%CITATION = ARXIV:1612.05991;%%

  \bibitem{Zumalacarregui:2013pma} 
  M.~Zumalac\'{a}rregui and J.~Garc\'{\i}a-Bellido,
  %``Transforming gravity: from derivative couplings to matter to second-order scalar-tensor theories beyond the Horndeski Lagrangian,''
  Phys.\ Rev.\ D {\bf 89}, 064046 (2014)
  [arXiv:1308.4685 [gr-qc]].
  %%CITATION = ARXIV:1308.4685;%%
  %56 citations counted in INSPIRE as of 24 sept. 2015
  

\bibitem{Gleyzes:2014dya} 
  J.~Gleyzes, D.~Langlois, F.~Piazza and F.~Vernizzi,
  %``Healthy theories beyond Horndeski,''
  Phys.\ Rev.\ Lett.\  {\bf 114}, no. 21, 211101 (2015)
  [arXiv:1404.6495 [hep-th]].
  %%CITATION = ARXIV:1404.6495;%%
  %65 citations counted in INSPIRE as of 23 Oct 2015
  

\bibitem{Gleyzes:2014qga} 
  J.~Gleyzes, D.~Langlois, F.~Piazza and F.~Vernizzi,
  %``Exploring gravitational theories beyond Horndeski,''
  JCAP {\bf 1502}, 018 (2015)
  [arXiv:1408.1952 [astro-ph.CO]].
  
  
\bibitem{Deffayet:2015qwa} 
C.~Deffayet, G.~Esposito-Farese and D.~A.~Steer,
%``Counting the degrees of freedom of generalized Galileons,''
Phys.\ Rev.\ D {\bf 92}, 084013 (2015)
%doi:10.1103/PhysRevD.92.084013
[arXiv:1506.01974 [gr-qc]].
%%CITATION = doi:10.1103/PhysRevD.92.084013;%%
%35 citations counted in INSPIRE as of 03 Aug 2016 

\bibitem{Langlois:2015cwa} 
  D.~Langlois and K.~Noui,
  %``Degenerate higher derivative theories beyond Horndeski: evading the Ostrogradski instability,''
  JCAP {\bf 1602}, no. 02, 034 (2016)
 % doi:10.1088/1475-7516/2016/02/034
  [arXiv:1510.06930 [gr-qc]].
  %%CITATION = doi:10.1088/1475-7516/2016/02/034;%%
  %12 citations counted in INSPIRE as of 25 Feb 2016
 

\bibitem{Langlois:2015skt} 
  D.~Langlois and K.~Noui,
  %``Hamiltonian analysis of higher derivative scalar-tensor theories,''
  JCAP {\bf 1607}, no. 07, 016 (2016)
 % doi:10.1088/1475-7516/2016/07/016
  [arXiv:1512.06820 [gr-qc]].
  %%CITATION = doi:10.1088/1475-7516/2016/07/016;%%
  %19 citations counted in INSPIRE as of 05 Aug 2016
  
\bibitem{Crisostomi:2016tcp} 
  M.~Crisostomi, M.~Hull, K.~Koyama and G.~Tasinato,
  %``Horndeski: beyond, or not beyond?,''
  JCAP {\bf 1603}, no. 03, 038 (2016)
  %doi:10.1088/1475-7516/2016/03/038
  [arXiv:1601.04658 [hep-th]].
  %%CITATION = doi:10.1088/1475-7516/2016/03/038;%%
  %16 citations counted in INSPIRE as of 05 Aug 2016

 
\bibitem{Crisostomi:2016czh} 
  M.~Crisostomi, K.~Koyama and G.~Tasinato,
  %``Extended Scalar-Tensor Theories of Gravity,''
  JCAP {\bf 1604}, no. 04, 044 (2016)
  %doi:10.1088/1475-7516/2016/04/044
  [arXiv:1602.03119 [hep-th]].
  %%CITATION = doi:10.1088/1475-7516/2016/04/044;%%
  %12 citations counted in INSPIRE as of 05 Aug 2016
 
\bibitem{Achour:2016rkg} 
  J.~Ben Achour, D.~Langlois and K.~Noui,
  %``Degenerate higher order scalar-tensor theories beyond Horndeski and disformal transformations,''
  Phys.\ Rev.\ D {\bf 93}, no. 12, 124005 (2016)
  %doi:10.1103/PhysRevD.93.124005
  [arXiv:1602.08398 [gr-qc]].
  %%CITATION = doi:10.1103/PhysRevD.93.124005;%%
  %9 citations counted in INSPIRE as of 05 Aug 2016
  
  
\bibitem{deRham:2016wji} 
  C.~de Rham and A.~Matas,
  %``Ostrogradsky in Theories with Multiple Fields,''
  JCAP {\bf 1606}, no. 06, 041 (2016)
  %doi:10.1088/1475-7516/2016/06/041
  [arXiv:1604.08638 [hep-th]].
  %%CITATION = doi:10.1088/1475-7516/2016/06/041;%%
  %1 citations counted in INSPIRE as of 07 Jul 2016
  
\bibitem{Ezquiaga:2017ner} 
  J.~M.~Ezquiaga, J.~Garc\'{\i}a-Bellido and M.~Zumalac\'{a}rregui,
  %``Field redefinitions in theories beyond Einstein gravity using the language of differential forms,''
  arXiv:1701.05476 [hep-th].
  
\bibitem{BenAchour:2016fzp} 
  J.~Ben Achour, M.~Crisostomi, K.~Koyama, D.~Langlois, K.~Noui and G.~Tasinato,
  %``Degenerate higher order scalar-tensor theories beyond Horndeski up to cubic order,''
  JHEP {\bf 1612}, 100 (2016)
  %doi:10.1007/JHEP12(2016)100
  [arXiv:1608.08135 [hep-th]].
  
  \bibitem{Heisenberg:2016eld} 
  L.~Heisenberg, R.~Kase and S.~Tsujikawa,
  %``Beyond generalized Proca theories,''
  Phys.\ Lett.\ B {\bf 760}, 617 (2016)
  %doi:10.1016/j.physletb.2016.07.052
  [arXiv:1605.05565 [hep-th]].
  
  \bibitem{Kimura:2016rzw} 
  R.~Kimura, A.~Naruko and D.~Yoshida,
  %``Extended vector-tensor theories,''
  JCAP {\bf 1701}, no. 01, 002 (2017)
 % doi:10.1088/1475-7516/2017/01/002
  [arXiv:1608.07066 [gr-qc]].
  

%\cite{Motohashi:2016ftl}
\bibitem{Motohashi:2016ftl} 
  H.~Motohashi, K.~Noui, T.~Suyama, M.~Yamaguchi and D.~Langlois,
  %``Healthy degenerate theories with higher derivatives,''
  JCAP {\bf 1607}, 007 (2016)
  %doi:10.1088/1475-7516/2016/07/033
  [arXiv:1603.09355 [hep-th]].
  %%CITATION = doi:10.1088/1475-7516/2016/07/033;%%
  %4 citations counted in INSPIRE as of 29 Jul 2016
  
  
 %\cite{Klein:2016aiq}
\bibitem{Klein:2016aiq} 
  R.~Klein and D.~Roest,
  %``Exorcising the Ostrogradsky ghost in coupled systems,''
  JHEP {\bf 1607}, 130 (2016)
  %doi:10.1007/JHEP07(2016)130
  [arXiv:1604.01719 [hep-th]].
  %%CITATION = doi:10.1007/JHEP07(2016)130;%%
  %2 citations counted in INSPIRE as of 05 Aug 2016 
  
  \bibitem{Motohashi:2014opa} 
  H.~Motohashi and T.~Suyama,
  %``Third order equations of motion and the Ostrogradsky instability,''
  Phys.\ Rev.\ D {\bf 91}, no. 8, 085009 (2015)
  %doi:10.1103/PhysRevD.91.085009
  [arXiv:1411.3721 [physics.class-ph]].
  
  \bibitem{Henneaux:2010vx} 
  M.~Henneaux, A.~Kleinschmidt and G.~Lucena Gomez,
  %``Remarks on Gauge Invariance and First-Class Constraints,''
  arXiv:1004.3769 [hep-th].
  
  \bibitem{Comelli:2012vz} 
  D.~Comelli, M.~Crisostomi, F.~Nesti and L.~Pilo,
  %``Degrees of Freedom in Massive Gravity,''
  Phys.\ Rev.\ D {\bf 86}, 101502 (2012)
  %doi:10.1103/PhysRevD.86.101502
  [arXiv:1204.1027 [hep-th]].
  
  
  %\cite{Deffayet:2010zh}
  \bibitem{Deffayet:2010zh}
  C.~Deffayet, S.~Deser and G.~Esposito-Farese,
  %``Arbitrary $p$-form Galileons,''
  Phys.\ Rev.\ D {\bf 82} (2010) 061501
 % doi:10.1103/PhysRevD.82.061501
  [arXiv:1007.5278 [gr-qc]].
  %%CITATION = doi:10.1103/PhysRevD.82.061501;%%
  %138 citations counted in INSPIRE as of 27 Feb 2017

%\cite{Padilla:2010de}
\bibitem{Padilla:2010de}
A.~Padilla, P.~M.~Saffin and S.~Y.~Zhou,
%``Bi-galileon theory I: Motivation and formulation,''
JHEP {\bf 1012} (2010) 031
%doi:10.1007/JHEP12(2010)031
[arXiv:1007.5424 [hep-th]].
%%CITATION = doi:10.1007/JHEP12(2010)031;%%
%108 citations counted in INSPIRE as of 27 Feb 2017

%\cite{Hinterbichler:2010xn}
\bibitem{Hinterbichler:2010xn}
K.~Hinterbichler, M.~Trodden and D.~Wesley,
%``Multi-field galileons and higher co-dimension branes,''
Phys.\ Rev.\ D {\bf 82} (2010) 124018
%doi:10.1103/PhysRevD.82.124018
[arXiv:1008.1305 [hep-th]].
%%CITATION = doi:10.1103/PhysRevD.82.124018;%%
%162 citations counted in INSPIRE as of 27 Feb 2017
  
  
  %\cite{Padilla:2012dx}
  \bibitem{Padilla:2012dx}
  A.~Padilla and V.~Sivanesan,
  %``Covariant multi-galileons and their generalisation,''
  JHEP {\bf 1304} (2013) 032
  %doi:10.1007/JHEP04(2013)032
  [arXiv:1210.4026 [gr-qc]].
  %%CITATION = doi:10.1007/JHEP04(2013)032;%%
  %43 citations counted in INSPIRE as of 27 Feb 2017  
  
  
 
  \bibitem{Allys:2016hfl}
  E.~Allys,
  %``New terms for scalar multi-galileon models, application to SO(N) and SU(N) group representations,''
  arXiv:1612.01972 [hep-th].
  %%CITATION = ARXIV:1612.01972;%%
  %1 citations counted in INSPIRE as of 27 Feb 2017
  
    %\cite{Sivanesan:2013tba}
    \bibitem{Sivanesan:2013tba}
    V.~Sivanesan,
    %``Generalized multiple-scalar field theory in Minkowski space-time free of Ostrogradski ghosts,''
    Phys.\ Rev.\ D {\bf 90} (2014) no.10,  104006
    %doi:10.1103/PhysRevD.90.104006
    [arXiv:1307.8081 [gr-qc]].
    %%CITATION = doi:10.1103/PhysRevD.90.104006;%%
    %22 citations counted in INSPIRE as of 27 Feb 2017
  
  \bibitem{Fairlie:1994in} 
  D.~B.~Fairlie and A.~N.~Leznov,
  %``General solutions of the Monge-Ampere equation in n-dimensional space,''
  J.\ Geom.\ Phys.\  {\bf 16}, 385 (1995)
 % doi:10.1016/0393-0440(94)00035-3
  [hep-th/9403134].
  
  \bibitem{Comelli:2013txa} 
  D.~Comelli, F.~Nesti and L.~Pilo,
  %``Massive gravity: a General Analysis,''
  JHEP {\bf 1307}, 161 (2013)
  %doi:10.1007/JHEP07(2013)161
  [arXiv:1305.0236 [hep-th]].
  
  
  
  
  
%\cite{Gracia:1988xp}
  \bibitem{DOFcount1}
  X.~Gracia and J.~M.~Pons,
  %``Gauge Generators, Dirac's Conjecture and Degrees of Freedom for Constrained Systems,''
  Annals Phys.\  {\bf 187} (1988) 355.
  %doi:10.1016/0003-4916(88)90153-4
  %%CITATION = doi:10.1016/0003-4916(88)90153-4;%%
  %39 citations counted in INSPIRE as of 06 Apr 2016
  
  %\cite{Pons:1986zg}
  \bibitem{DOFcount2}
  J.~M.~Pons,
  %``New Relations Between Hamiltonian and Lagrangian Constraints,''
  J.\ Phys.\ A {\bf 21} (1988) 2705.
  %doi:10.1088/0305-4470/21/12/014
  %%CITATION = doi:10.1088/0305-4470/21/12/014;%%
  %20 citations counted in INSPIRE as of 06 Apr 2016
  
  %\cite{Henneaux:1990au}
  \bibitem{DOFcount3}
  M.~Henneaux, C.~Teitelboim and J.~Zanelli,
  %``Gauge Invariance and Degree of Freedom Count,''
  Nucl.\ Phys.\ B {\bf 332} (1990) 169.
  %doi:10.1016/0550-3213(90)90034-B
  %%CITATION = doi:10.1016/0550-3213(90)90034-B;%%
  %129 citations counted in INSPIRE as of 06 Apr 2016
  
  
  %\cite{Diaz:2014yua}
  \bibitem{DOFcount4}
  B.~Díaz, D.~Higuita and M.~Montesinos,
  %``Lagrangian approach to the physical degree of freedom count,''
  J.\ Math.\ Phys.\  {\bf 55} (2014) 122901
  %doi:10.1063/1.4903183
  [arXiv:1406.1156 [hep-th]].
  %%CITATION = doi:10.1063/1.4903183;%%
  %1 citations counted in INSPIRE as of 06 Apr 2016 
  
  \bibitem{Dirac} 
  P. Dirac, {\it Lectures on Quantum Mechanics}.
  Dover Books on Physics. Dover Publications, 2001.

 
  
  %\cite{deRham:2013hsa}
  \bibitem{deRham:2013hsa}
  C.~de Rham, M.~Fasiello and A.~J.~Tolley,
  %``Galileon Duality,''
  Phys.\ Lett.\ B {\bf 733} (2014) 46
  %doi:10.1016/j.physletb.2014.03.061
  [arXiv:1308.2702 [hep-th]].
  %%CITATION = doi:10.1016/j.physletb.2014.03.061;%%
  %45 citations counted in INSPIRE as of 01 Mar 2017
  
  %\cite{deRham:2014lqa}
  \bibitem{deRham:2014lqa}
  C.~De Rham, L.~Keltner and A.~J.~Tolley,
  %``Generalized galileon duality,''
  Phys.\ Rev.\ D {\bf 90} (2014) no.2,  024050
 % doi:10.1103/PhysRevD.90.024050
  [arXiv:1403.3690 [hep-th]].
  %%CITATION = doi:10.1103/PhysRevD.90.024050;%%
  %36 citations counted in INSPIRE as of 01 Mar 2017
  

\end{thebibliography}
\end{document}